# The G-Ball, a New Icon for Codon Symmetry and the Genetic Code[1]

by Mark White, MD
Copyright Rafiki, Inc. 2007.

> "The most exciting phrase to hear in science, the one that heralds new discoveries, is not 'Eureka!' but 'That's funny...' "
>
> Isaac Asimov

**Abstract**: A codon table is a useful tool for mapping codons to amino acids as they have been assigned by nature. It has become a scientific icon because of the way it embodies our understanding of this natural process and the way it immediately communicates this understanding. However, advancements in molecular biology over the past several decades must lead to a realization that our basic understanding of genetic translation is fundamentally flawed and incomplete, and, therefore, our icon is inadequate. A better understanding of symmetry and an appreciation for the essential role it has played in codon formation will improve our understanding of nature's coding processes. Incorporation of this symmetry into our icon will facilitate that improvement.

## Introduction to an Icon

The standard codon table has failed as an icon of the genetic code. It fails to capture the basic structure and function of nature's real code of protein synthesis. The codon table belies none of the logic that led to the existence of this simple table, the data now filling it, or its true function in nature. It is a tiny and demonstrably incomplete set of data that is merely arranged by the arbitrary structure of the table itself. It does this in such a way as to merely support and amplify a false model of collapsed molecular information and thereby fails to predict or explain the ultimate formation of any protein structure. Quite simply, it is a failed icon that perfectly represents the flawed features of a failed model of protein synthesis. Therefore, the standard codon table as an icon must be seriously analyzed, ultimately rejected, and then replaced with something more robust. Our thinking about genetic translation in general must begin to change in radical ways, and the icon we chose here will strongly inform our thinking as it changes.

This paper introduces the essential features and methods for constructing a new, multi-dimensional, perfectly symmetrical icon called the G-ball, one that still fails to embody the entire code in question, as all codon maps must, but one that is more reflective of the logical structure behind the code itself. This new structure has many epistemic implications. After all, the data within any codon table might somehow be organized to reflect the larger reality of a complex, symmetrical, multi-dimensional matrix of information relating nucleotides to proteins. And it is true that the genetic code itself is not a simple, linear substitution cipher as the standard table graphically portrays it, but the icon we use to illustrate this limited molecular information can still reflect at least some of the organizational properties of the translation system itself. The problem of graphically illustrating codons should now be seen as somewhat analogous to searching for a data structure similar in pedagogic function to the periodic table of elements that is so useful in chemistry. Data form can inform data.

Just as the structure and natural symmetry of DNA's double helix informs our thinking about genomes[2], the structure and natural symmetry of codons should inform our thinking about the genetic code. An enlightened view of molecular information reveals that the double helix and the genetic code actually share symmetry in many different and important ways. In other words, fundamental laws of universal symmetry appear to organize the miraculous system of molecular translation involved in the genetic code. These laws appear to have acted on complex sets of self-organized molecules over vast periods of time to ultimately give us this wonderful view of life that we find in virtually every living cell today.

The term "the genetic code" is merely a linguistic icon that we use to signify our model of the natural processes behind protein synthesis. The name, the model, and our organization of these particular molecular data subsets are nothing but a human metaphor[3] of nature's molecular metaphor. In truth, this code should rightly be called the protein code. However, our current view, its language and icons are now deeply entrenched, and they have been entirely derived from a false "one-dimensional" model of the genetic code. In our minds today only one dimension of "co-linear" molecular information can be translated by this code. Information passes from a nucleotide sequence to an amino acid sequence in one form only, and then this single dimension of information supposedly goes on to mysteriously define a fully formed protein[4]. It is this overly simplistic view of the genetic code that now serves as a formal and rigid definition of "molecular information." The same must also then be true for molecular information within larger models that must use the genetic code as a paradigm or conceptual point of initial reference. The one and only dimension involved in this model is easily and quickly found in the codon table, and it exists only as a simple relationship between three nucleotides, independent of all other context. These nucleotides must also always be selected from a set of only four possible nucleotides. This is, in fact, what rigidly defines a codon today, and this is what dictates the limits of our understanding. The simplistic genetic code model and its inseparable visual icon, the codon table, therefore, provide us only with a cipher to determine translated sequences of amino acids but not actually the translation of whole proteins in any real sense. In other words, codons today literally mean amino acids and sequences of amino acids literally mean proteins in this model.

So, the codon table is for all practical purposes a comprehensive representation of the entire genetic code today. But this is not logically an acceptable model. Silent mutations change folded proteins[5,6] and this empiric evidence – in addition to common sense and mountains of other evidence - clearly demonstrates at least more than one dimension of information acting in translation. This has once and for all convincingly proven the central premise of one-dimensionality to be utterly false. So, science is now adrift without an ideological rudder in this area of thought.

The codon table, although technically a "linear" data structure, is usually arranged in a two-dimensional grid of data that must always treat the data asymmetrically. It subjectively weights the data via choices that must be made based on asymmetry. It presents this data in a compressed and graphically convenient format, to be sure, yet it is still contained in only a partially compressed format. So, any specific arrangement of this type and of this data must always be largely subjective in this way. Therefore, any patterns that appear in it are also subjective to a large degree. The visible data patterns become largely a result of the patterns of table construction. However, a more symmetric, multi-dimensional yet maximally compressed and virtually objective arrangement of this same data will be more informative toward our knowledge of the genetic code, to be sure.

**What are Codons?**

The codon table is a map. It maps the set of codons to the set of amino acids. Presumably, the function of the genetic code then is to translate codons into amino acids, and the map of this can be called a graph of this function. The two sets in question now are undeniably codons and amino acids. Set A is codons and set B is amino acids, and yet there is still a certain amount of noticeable symmetry between them. When we talk of functions between sets we say that set B is the image of set A, or set A projects onto set

B. We might also rightly say that set A is cause and set B is effect. For every cause in nature there generally must be a single effect, yet many causes can have the same effect, and so there must be at least as many causes as there are effects. This is a simple idea that is generally valid and widely acknowledged, and one that Joe Rosen[7] describes as a universal symmetry principle. In other words, the universe is a symmetrical place with respect to cause and effect, and this symmetry principle holds that effects must be at least as symmetrical as causes. This could be seen as a basic axiom of science that should not generally be violated. We might uniformly reject any notion that it is being violated in nature. If a theory violates the symmetry principle, then we can strongly intuit that the theory is false in some fundamental way. When silent mutations alter folded proteins, codons cannot literally mean amino acids during translation. Therefore, the standard codon table and the simple concept of "linearity" that it represents are in clear violation of the universal symmetry principle. Furthermore, the standard codon table cannot then stand as a satisfactory icon of the genetic code. It's just that simple.

To begin to partially rectify this nasty situation, we must seriously address a formal process of defining cause and effect in molecular translations, and then start searching for potentially adequate pairings of the two. Effects should not exceed causes. We must begin by defining codons in a more abstract way, and doing this somewhat formally so that we can begin to identify the actual molecular set and pair it to other sets involved in translation. Because symmetry clearly plays a major role in this natural system of translation, symmetry is a good place to start when defining the individual components within a model of this system. We will start here with the symmetry of nucleotides and then add in the natural symmetry of codons. We will then see how this can potentially map to the now apparent yet still mysterious symmetry found in the standard set of amino acids. Although codons cannot mean amino acids in translation, they can still share common symmetries.

A group is a formal concept in mathematics, and group theory is the mathematician's preferred language of symmetry. Just as a number is a measure of quantity, a group is a measure of symmetry. This basic notion and the formal language built around it give us useful tools for defining and describing a system of molecular translations. Symmetry itself is an entirely abstract concept. It exists in many different ways that aren't always formally recognized, yet we generally know it when we see it. The ancient Greeks, who were noted for their appreciation of symmetry, saw it purely as a form of analogy. Symmetry is the fundamental invariance within a relationship of one thing to another. This is still a useful way to think about symmetry – as comparison - but we will delve into more formal uses here. To wit, a mathematical group is a well-defined set of transformations that create consistent compositions within the set of transformations. This basically means that objects in a set can be acted upon by symmetry transformations and merely generate other objects in the same set. I will not go into group theory in much depth or sophistication here, but I must briefly use it as a tool to advance our general language and further our illustrations of nucleotides and codons. These general yet somewhat formal ideas about symmetry will greatly help inform our thinking about any new icon and model of the genetic code.

The set of positive integers plus zero when acted upon by simple addition, for instance, is perhaps the most common example given of a symmetrical set of numbers. Any two integers when added together merely produce another integer. All integers are symmetrical with respect to all transformations of addition within the set. Addition is the symmetry and integers are a set of numbers that demonstrate it. But sets of elements of common geometry can be more tangible and robust examples of symmetry. The six faces, eight points and twelve edges of a cube illustrate perhaps a more useful example of spatial symmetry. Integers are an example of linear symmetry but a cube is an example of spatial symmetry. However, symmetry itself is merely an abstract form of transformation, and it can manifest in any form.

The elements of a cube are more concrete visual demonstrations of a symmetry group than is the set of integers. We can visualize a full set of transformations of a cube in space that include all rotations and mirror symmetries that always leave the cube in a final form that is indistinguishable from its initial form. The real spatial elements of any cube can be easily rotated and reflected through space yet leave the cube itself fundamentally unaltered via the cube's inherent symmetry properties. It is the abstract symmetry properties of the cube that define the group of transformations. The cube itself is merely a realization of this group, existing only as a real set of points, faces and edges in real space. It is easy to confuse the cube with the symmetry group that it represents. Symmetry defines actual sets and these sets can clearly represent that symmetry. The cube is a real set of elements. The symmetry group of the cube is a set of transformations that can be performed on the cube. The same is always true of molecules; therefore, we will want to start our definitions of molecular sets with symmetry and not, as is conventionally done, with actual sets of molecules.

Spatial symmetry has proven quite useful in modeling and understanding the self-assembly processes in many different inorganic molecular systems in nature, such as a salt crystal. It is also helpful with many organic examples, such as virus particles. However, geometric symmetry can be surprisingly useful for visualization and understanding of other molecular sequence symmetries. This will allow us to conceptually merge the physical self-assembly principles behind sequence and structure in DNA and its logical involvement in protein synthesis. Symmetry is the molecular unifier. Symmetry is the glue that binds molecular information of all forms.

As it turns out, much to everyone's surprise, virtually anything might represent a symmetry group, like, for instance, the set of solutions to specific types of algebraic equations[8]. A set of things reflects a formal mathematical group if only their transformations can satisfy four abstract criteria, which are: associativity, possess an identity element, possess an inverse element, and demonstrate closure. That's it. The details are less difficult than they may seem, and an explanation can be found elsewhere[9]. However, we can easily use this general notion here to conceptualize the symmetry in the nucleotide sets involved in the structure of DNA's double helix, and then further use it in our definition of codons. Although, and this is an unexpectedly tricky point with respect to real world biochemistry, deciding in a practical sense on a precise definition of a codon is less obvious when put into this formal yet more abstract setting. How exactly should a codon be defined in nature? The hidden subtleties of this definition, it turns out, lie principally behind today's widespread confusion.  Any chosen parameters in the definition of a codon will significantly impact the size and overall structure of any codon set, and by extension all of the other molecular sets with which it is symmetrically related during translation of any molecular code.

Conventionally, it has been immediately noted that there are exactly four nucleotides in DNA, and that these four are merely combined in sequences of three consecutive nucleotides to make up a set of sixty-four codons ($4^3$ or 4 X 4 X 4 = 64). This is backwards thinking in this context, and it seems to simplify the definition of the symmetry of any set merely to an examination of its overall size.  This is, in fact, a less-than-adequate definition of codons for a variety of now obvious reasons.  Perhaps the best reason is that there are several more than four nucleotides that participate in the real-world system of translation that we have uncovered in nature.  The real-world image of any set of codons is a set of anticodons.  Codons directly "mean" anticodons in nature, and the set of anticodons is still undefined and perhaps largely unknown.  So, when that fifth nucleotide shows up in translation, as it inevitably does in nature (a bit like the fifth Beatle) how do we then fit it into our definition of codons, their image, and the resulting new sizes of their molecular sets?  What are the appropriate mappings when the image exceeds the original set?  In a system with at least five nucleotides, what now is a codon? This is why a more general description is appropriate here and why it can serve us well as an example of much needed abstraction in this area as we drill down on the real-world structures built upon useful invariant natural symmetries.  These symmetry principles are clearly evident in this molecular information system that we call the genetic code, and remarkably they can be reflected in a proper icon for it.

Codons are made of nucleotides and nucleotides demonstrate a fundamental symmetry.  We can begin here to appreciate a more general description of codons by first looking at individual nucleotides and their inverse nucleotide pairs.  This can be done abstractly with a simple schematic of a short, generic DNA sequence of nine base-pairs.

Figure 1.

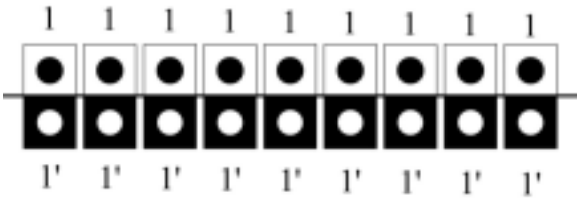

This schematic illustrates the known fact that the double helix of DNA is comprised of a sequence of bases, and for each base in the sequence (1) there will exist at least one complement to that base (1'). In the language of groups, a base complement can be considered the inverse of the base. If we imagine two nucleotides pairing in nature, we can imagine a point of inversion between the two bases. The primary logical structure of DNA's double helix is built only from the concept of two complementary strands considered one element at a time. The translation of DNA into more DNA, also known as DNA replication, occurs one base at a time; therefore, there is really no inherent direction to the sequence of DNA with respect to the logic of translation into more DNA. It can be equally well translated in one direction as the other, and so it is. Complement formation is the only translation operation performed on the set of bases, and so it can easily be shown that they constitute a simple group with respect to the logic of DNA replication.

Table 1.

| **E** | 1  | Identity |
|-------|----|----------|
| **i** | 1' | inverse  |

Table 2.

|       | **E** | **i** |
|-------|-------|-------|
| **E** | E     | i     |
| **i** | i     | E     |

Besides nucleotide identity (identity is the trivial form of symmetry that is always included in every symmetry group) there is only one element in the set of single nucleotide transformations of DNA. This is analogous to logical inversion, and it is easy to show that this small transformation set forms a symmetry group. In this context we might imagine that a sequence of base pairs represents a "linear crystal" or a linear lattice of points[10,11]. The unit cell of this lattice is the displacement of a single point in either direction. If the sequence is infinite, then the symmetry is perfect. If the sequence is finite, as it always is in nature, then displacements cannot be performed equally on every point, so we say that it shows approximate symmetry. Most systems in nature can only show approximate symmetry because of obvious physical boundaries and obvious symmetry breaking. However, the natural symmetry of this group is abstractly independent of the number of bases in any particular set. There could be 1, 2, 3, 4, 10, 1024, or any given number of bases in a set with this symmetry, and they could be equally divided into exclusive pairs but need not be. When specific bases are chosen for a particular set we can say that the symmetry is broken. There are some ways to break symmetry that are more symmetrical than others. The fact that nature happened to give us a set of four bases – A, C, G, T, two exclusive pairs of bases - is significant. It is a dual binary, or literally a two-bit system. The empiric fact that nature has broken the symmetry of nucleotides in exactly this way conveniently allows us to now objectify and visualize this particular set of four bases. We can do so by using the perfect arrangement of dual faces on an octahedron. We can graphically illustrate this specific set of four bases and their complementary symmetry by putting them on the faces of an octahedron. This allows us to better visualize the symmetry of this special case with respect to real-world translation operations of DNA into more DNA. Symmetry is abstract but its logic can be made visible by real sets of objects that share symmetry.

Figure 2

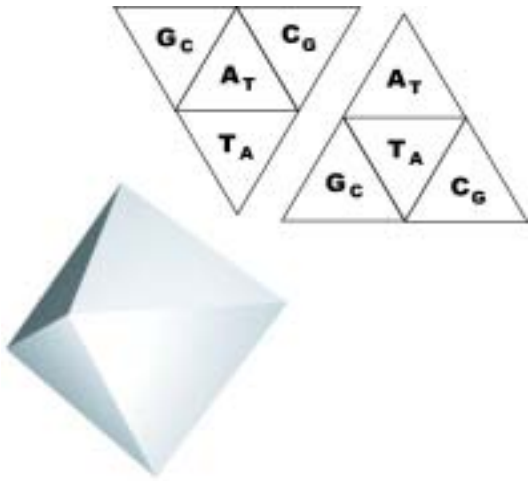

Each face of an octahedron can be labeled with a base and a subscript, called a McNeil subscript,[12] and the subscript in this special case will tell us the complement, which also happens to be the base on the opposite face. The centroid of the octahedron acts as a point of inversion for its eight faces with respect to the set of four possible base pairs. This is merely one example of how elements of common geometry can help us visualize the abstract symmetry in a specific set of molecules. (The same mapping of this particular information also has a mirror version, but it is irrelevant to the discussion here.)

By comparison, the translation of DNA into protein, or operations of "the genetic code" when compared to this simple case of translating DNA into DNA introduces but a single new logical feature to the translation system at this basic level. Instead of merely operating on one base at a time, the bases are now "read" three bases at a time. If it is again seen as a "linear crystal" then the unit cell of the lattice becomes a set of three points that is displaced in a single direction. Consecutive nucleotides become consecutive codons. Independent of the actual number of bases in the set, we can again schematically illustrate this system.

Figure 3

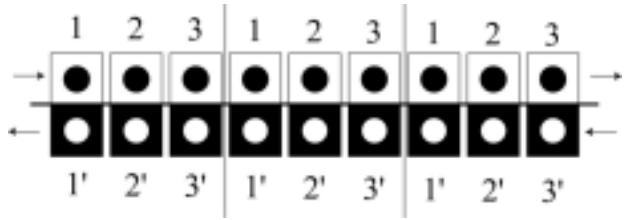

In going from a reading frame of one base to a reading frame of three bases we have introduced a logical reading direction.  We have created ordered sets of three nucleotides.  We empirically know that DNA is structured such that there is a physical difference between the "beginning" and "end" of any DNA sequence, and this difference is inverted in the complement sequence.  The double helix of DNA contains a "coding strand" of nucleotides that has a logical reading direction, and a "non-coding strand" where the nucleotides and the reading direction are inverted.  DNA is a natural two-for-one deal with respect to nucleotide sequences.  Codons, for all intents and purposes, travel in pairs.  This schematic gives us a picture of the standard orientation or a proper "reading frame" within which we can now define codons.  A codon is now simply an ordered set of three nucleotides.  Figure 3 labels the bases 1, 2 and 3, and their complements 1', 2' and 3'.  This has nothing to do with the specific identity of the base in any particular sequence, but rather only the position of a base within a given reading frame.  So again, the symmetry of this translation system can remain completely independent of the actual number of bases in any set.  However, the number of actual elements between mappings of any two sets of this symmetry needs not be the exact same.

This brings us to an important observation: codons are not "real" in the normal sense of the word.  In other words, we cannot find a codon existing independently anywhere in nature.  They are molecular subsets that can never exist as a sovereign molecule in the way we typically define a molecule.  Three nucleotides do not represent a codon independent of context.  Codons are manifestations of individual nucleotides, specific sequences of nucleotides, and the ordering of sets within larger sets of those nucleotides, existing only as the *relationships* between nucleotides. Codons define reading frames and reading frames define codons. Every codon only exists relative to other codons. Since these sets are ordered, and since these sequences commonly change, the sets are also commonly re-ordered. It is the ordering and reordering of nucleotides that defines codons and their inherent symmetry. Codons are not real and they are not static. Codons exist only as a dynamic relationship between specific nucleotides in sequence, and that relationship is then dynamically related to other molecular parameters during the process of molecular translation.

To formally define the symmetry group of codons we must identify all transformations of three ordered nucleotides. This is not too difficult because it is merely a common set of sequence permutations, and there are only six ways to permute a set of three sequential elements:

123, 231, 312, 132, 213, 321

Cayley's theorem tells us that every group is isomorphic to a subgroup of a group of permutations; therefore, any physical object with symmetry that matches the permutations of a codon can be used to illustrate codons. The obvious way to illustrate this simple symmetry group - known formally as dihedral symmetry $D_3$ - is with a triangle of points labeled 1, 2 and 3.

Figure 4.

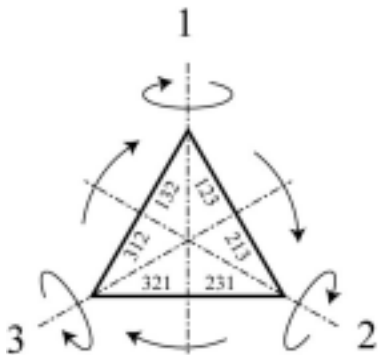

A triangle can be rotated three times around an axis perpendicular to it. It can also be mirror reflected across any bisecting line. However, the three mirror planes have the same practical effect as a two-fold rotation on this axis. As illustrated here, we can more easily find all of these permutations within similar spaces on the triangle if we merely use a simple reading convention of points in both directions around the triangle. These symmetries are formally denoted by common convention and notation as follows:

Table 3.

| **E** | 123 | Identity |
| **r** | 231 | rotate 120 degrees |
| **$r^2$** | 312 | rotate 240 degrees |
| **m** | 132 | Mirror |
| **mr** | 231 | rotate 120 degrees and mirror |
| **$mr^2$** | 321 | rotate 240 degrees and mirror |

The multiplication table that proves this set of transformations is a symmetry group is as follows:

Table 4.

|  | **E** | **r** | **$r^2$** | **m** | **mr** | **$mr^2$** |
|---|---|---|---|---|---|---|
| **E** | E | r | $r^2$ | m | mr | $mr^2$ |
| **r** | r | $r^2$ | E | $mr^2$ | m | mr |
| **$r^2$** | $r^2$ | E | r | mr | $mr^2$ | m |
| **m** | m | mr | $mr^2$ | E | r | $r^2$ |
| **mr** | Mr | $mr^2$ | m | $r^2$ | E | r |
| **$mr^2$** | $mr^2$ | m | mr | r | $r^2$ | E |

There is nothing particularly complex or illogical about this view of codons, but this view should change the entire way we perceive codons. They are sets of elements related to each other by symmetry. We have now defined DNA's symmetry as nucleotide inversions in base pairs. We have also defined codon symmetry as being isometric with an equilateral triangle. We have identified both symmetry groups, and we can now combine the two symmetries and produce mappings for codons and their inversions on "non-coding" strands of DNA. The two-for-one nature of DNA means that codons must always travel in pairs.

Figure 5.

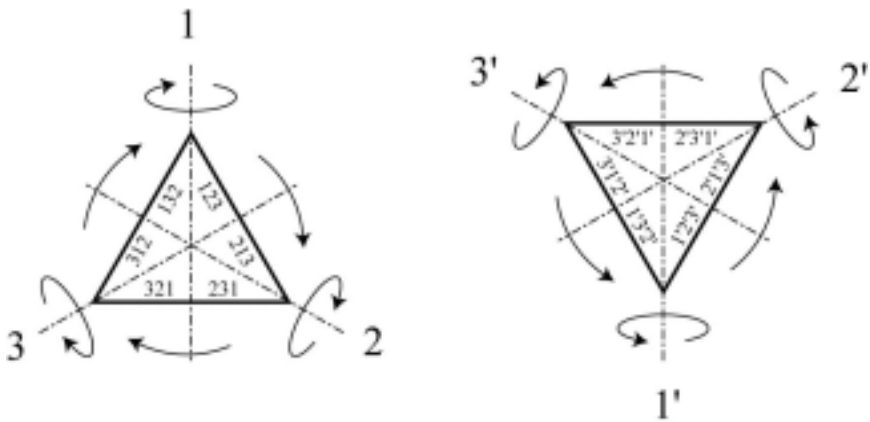

These symmetry groups are independent of the actual number of nucleotides and say nothing of whether they organize neatly into mutually complementary pairs as is seen in nature. They are purely manifestations of sequences and the inherent symmetry of their common transformations. The total number of actual codons in any set will be determined by a variety of factors. However, the groups themselves are now independent of the size of any particular set that may use them.

The logical independence of group and set size can now be better appreciated in the real world of biochemical data. Codons are translated into anticodons and not amino acids per se. Codons literally mean anticodons not amino acids. There is convincing evidence that more nucleotides exist in the set of anticodons than there are in the set of codons, so logically there are potentially more anticodons than codons. This is a simple mathematical relationship but it is commonly misunderstood in a bizarre way, and so one frequently hears the erroneous idea that there are fewer anticodons than there are codons. This is logically false. However, the true number of possible anticodons is independent of the actual number of molecules that possess them in nature. Nature has choices here, and we can expect her to take good advantage of them. The plain fact is, codons and anticodons share the same symmetry group, yet they are distinct molecular sets with different numbers of elements. The set of actual codons is translated into a potentially larger – or smaller - set of actual anticodons in nature. If the set of codons is not large enough to account for its image, then we simply must begin to consider the set of codon combinations in any effort to find the proper larger set. However, the mapping of one into the other depends at first upon a definition of the sets, preferably based on the structure and inherent symmetry and not solely on the actual size of the two sets.

This kind of basic abstraction begins to cut the wheat from the chaff and clear a path to a better understanding of the particular molecular information systems in question. It provides clues to how they could have possibly evolved, and how they might operate in nature. Symmetry plays a primary and not a secondary role in this context. The system itself is founded on natural symmetries. Furthermore, this same pattern can be traced up and down the complex hierarchy of this particular molecular translation system, which is actually a stunningly complex system – not a simple one. There are many sets, many relationships, and many different forms of molecular information involved. It is obviously more difficult to visualize this system and therefore comprehend the implications of this as we begin to add real data in moving forward toward our construction of a more appropriate icon of the genetic code.

Now that we have the general pattern of codon symmetry and have proven that they actually do form a symmetry group, we can begin to build tools to help us better visualize the common set of codons. We will then begin to recognize that it is the basic structure of the symmetry group that has significantly influenced the formation of the system of molecular translation that we call life, and not vice-versa.

## A Better Visualization of the Codon Symmetry Pattern

It is now apparent that the codon group and DNA are isomorphic with a set of dual triangles per Cayley's theorem. Perhaps not as apparent is the fact that the first triangle is merely combined with the group of DNA complements being translated into more DNA to generate the second dual triangle. Each strand of DNA is related to the other by its complements. DNA is a two-for-one deal of inverse strands. Notably, this is not the first time that something like this simple visualization technique has been done, at least in part. In 1957 the brilliant and colorful physicist, George Gamow, turned his attention to the nascent codon map and produced a similar, albeit a less robust model, one that he called the compact triangle code[13].

Figure 6.

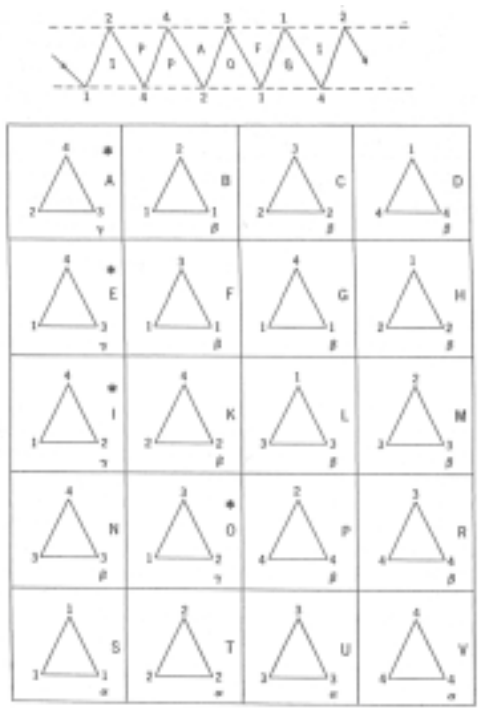

FIGURE 14. George Gamow, Alexander Rich, and Martynas Yčas, "The Problem of Information Transfer from Nucleic Acids to Proteins," *Advances in Biological and Medical Physics* 4 (New York: Academic Press, 1956), pp. 48–49. Bottom, twenty possible triads of triangular code; top, schematic diagram of triangular code. Reprinted by permission.

The good Dr. Gamow was on the right track but quite unfortunately fell well short of the conceptual mark on several counts. He was obviously hampered by a lack of data and what now appears to be a misunderstanding of the actual physical mechanism of translation. After all, he knew nothing of mRNA, tRNA and anticodons when he proposed his model. Then as now, a mapping of codons to amino acids is a mapping of the wrong sets of molecules with respect to the real-world functions of the genetic code. We continue to repeat Gamow's basic mistake today, yet this false perception is precisely what a codon table tells us to do.

First, Dr. Gamow assumed that his model should be based purely on an assumption of four nucleotides that can only form two sets of virtually exclusive base pairs. This is unfortunately still the accepted traditional approach to defining codons and it is specifically how he arrived at his model. Today's model always starts with DNA and builds upward, when a more enlightened view should start with codons and build upward and downward simultaneously. Second, he failed to consider the possibility that additional complementary triangles might actually somehow provide further insight of the overall pattern. In other words, he considered only twenty triangles when in fact there could be at least forty, possibly many more triangles, even within his own general scheme if made more abstract. Third, he failed to integrate his triangles into a comprehensive symmetry relationship. In fact, the basis of his model retrospectively seems to be predicated on the notion that global codon assignments will somehow reflect a symmetry minimum instead of a symmetry maximum. This could also be stated in terms of amino acid symmetry. In other words, he believed that amino acids are the image of codons and therefore must have at least the same degree of symmetry as codons. This is false. Amino acids are not the image of codons and have empirically been demonstrated to not compress their abstract symmetry as he expected. Finally, he apparently failed to rigorously test his model, presumably on the assumption that it had failed with empiric mapping of the first two codons, a failure that remarkably extends throughout all of the codon assignments to perfection. However, Gamow's perfect failure can further inform our thinking in a delightful fashion today. There is utility in failure, especially so in perfect failure.

As we begin to break the perfect symmetry of a codon, we should realize that there are only three general ways to break it.

1=2=3, 1=2≠3, 1≠2≠3

In other words, with respect to symmetry and symmetry breaking, there are three classes of codon. Gamow realized this and named them α, β and ᴦ, but I did not know this when I renamed them class I, II and III. I prefer my scheme and so I will continue to use it. We can add color to our original triangles and immediately see the logical difference between the three codon classes.

Figure 7.

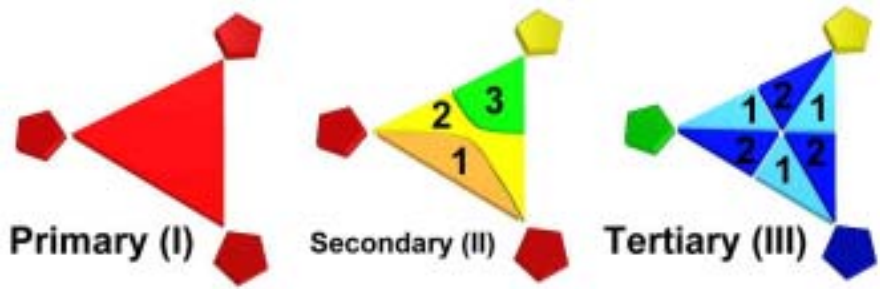

Within each class there are also different combinations of permutations that are equivalent, which I call codon types. In class I, all of the permutations are equivalent, so there is only one type of Class I codon. In class II they form three pairs of equivalent permutations, or three distinct types of codon, and in class III there are two sets of loosely related roto-isomers. Class III actually represents six non-equivalent permutations. Independent of the actual nucleotides in any set of codons, all codons share symmetry, and every specific instance of any codon can maintain more or less of this abstract symmetry. However, every set of actual codons can be organized globally around their relative symmetries. Gamow predicted that every codon would maintain its perfect symmetry with respect to every amino acid within each class. In other words, he predicted that every triangle would be assigned only one amino acid. It was an asymmetrical way to break global symmetry. This was perfectly wrong, and for reasons that are not obvious within any standard model. Amino acids do not perfectly maintain codon symmetry they perfectly break it. What we have heretofore failed to realize is that the relationship between one codon and another is always a part of the actual meaning of any codon. Symmetry is comparison and comparison is meaning in the world of molecular information. Symmetry organizes meaning within molecular information systems. Symmetry and symmetry breaking are always the first principles of molecular information.

As we start to break the perfect symmetry of codons, replacing them with the approximate symmetry of actual nucleotide sets, we can now see that there are several ways to actually break this symmetry in the real world of molecules. Had nature chosen Gamow's strategy, the system would have been efficient in one sense, but horribly inefficient in a more important way. It would mean that every codon would contain a minimum of information with respect to its own symmetry. Gamow was imagining a less robust system of translation, and it is hard to imagine a practical use for this kind of symmetry breaking now, given our current knowledge of how the actual translation system works. It does, however, make sense at the level of understanding that Gamow had of the system when he made his ingenious proposal. After all, Gamow was the only one at the time with the right idea, but he unfortunately proposed a perfectly incorrect solution to the problem. The question now becomes: Is there a way to perfectly break this global symmetry with nucleotides and amino acids? The answer, it turns out, is yes. To see this, we will require a far more enlightened view of codons and several additional tools of visualization.

In the same way that I objectified DNA symmetry with respect to replication transformations I will now use elements of common solid geometry to objectify and visualize the set of actual codons and thereby build the G-ball. Because the illustration quickly becomes heavy with numerous visual elements, I will again introduce colors as a way to quickly distinguish visually the various elements. Starting with the four nucleotides of DNA, we can objectify them as a single tetrahedron with a different base at each vertex. (Henceforth I prefer the RNA base U to the DNA base T.)

Figure 8.

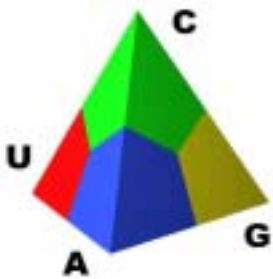

We can now easily see that four base poles create two dual axes in space predicated on their special known rules for base-pairing. One axis aligns the A:U poles and the other aligns the C:G poles. However, we still need a minimum of twelve base elements to generate all possible permutations for this specific translation system of nucleotide triplets; therefore, I will add a class I equilateral triangle representing three base elements perpendicular to each pole.

Figure 9.

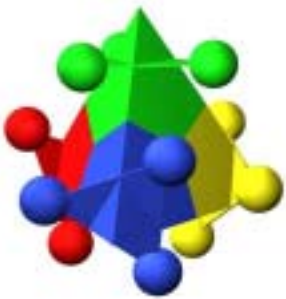

Conveniently, the points of these four triangles can be made to correspond perfectly with the face centers of a dodecahedron. Still more convenient is the fact that these points then generate sixteen additional equilateral triangles corresponding to the twenty triangular faces of an icosahedron, since the dodecahedron is a dual to the icosahedron.

Figure 10.

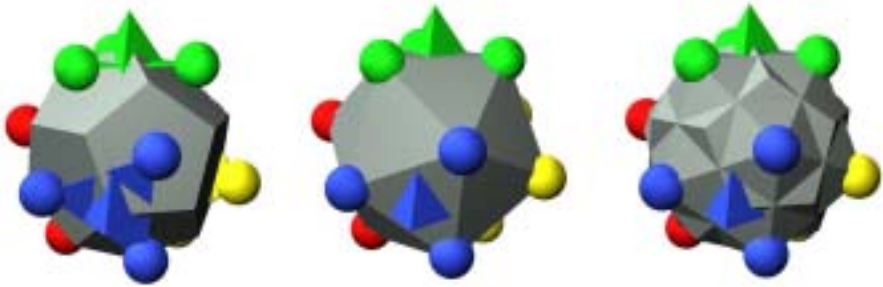

Happily, we have now generated all twenty equilateral triangles that Gamow included in his model. Still more happily, since this specific case involves only two complementary pairs of nucleotides, we have also generated the twenty complementary triangles as well. In fact, we have generated every possible permutation in the table that generally reflects the global symmetry of codons and codon complements - but this is true only for this specific set of molecules. This is a surprisingly simple procedure that should be viewed as significant. The set of DNA nucleotides does not give us the symmetry of codons but it does perfectly break the global symmetry of all codons. Life chose this pattern for a very good reason.

Furthermore, since this specific case involves only four nucleotides, the equivalent permutations of every triangle can be combined and related to all other permutations. We end up with only sixty-four unique permutations and not the 120 or 240 that we might expect from a more general case. In other words, we have used the dodecahedron and this specific set of four bases to quickly boil the pattern down to twenty sets of triangles with only sixty-four distinct permutations instead of built up to these numbers from the more simplistic first principles of our standard model.

We can further organize all of the codon types into four distinct super-sets based on the dominant base poles that contribute most strongly to each individual permutation. Within each pole we can sub-divide sets of permutations based entirely on single rotation symmetry, which I have called a multiplet of four codons. A multiplet is a collection of four permutations derived from common bases at the first two positions of every codon. These are also called wobble groups or family boxes elsewhere in conventionally inferior tables. Regardless of their general name, there are now obviously two basic types of multiplets, homogenous and heterogenous. The first one makes a circle in this mapping scheme and the other looks like a fish, at least it does to me. Each pole consists of three heterogeneous multiplets and one homogeneous multiplet. When combined into a coherent pattern of a dominant nucleotide pole, the four contiguous multiplets look to me like a flower. Every pole and multiplet has the same transformational symmetry patterns that we will visit a bit later.

Figure 11.

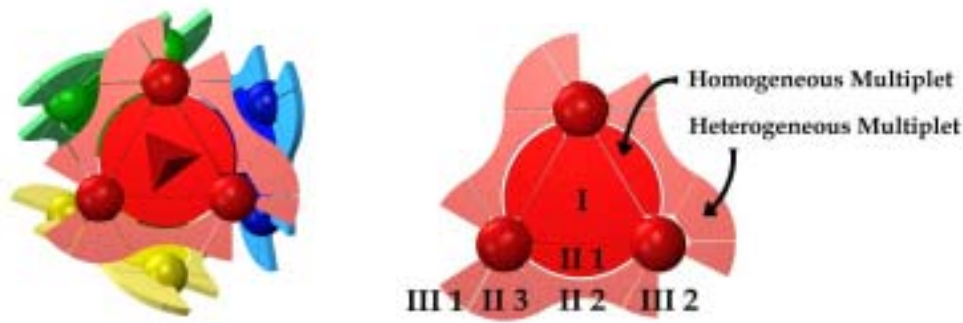

These visualization techniques merely represent graphical conventions based on common elements of geometry that are allowed only by the unique situation here that we are visualizing a set of two exclusively complementary sets of base pairs. If more bases are introduced, or if the pairing rules were to change, then these graphic techniques are perhaps no longer effective.  Under more complex circumstances, such as tRNA and anticodons, a similar, presumably a larger graphic structure could be constructed, but it will perhaps not be perfectly and comprehensively represented by the geometric symmetry of a single dodecahedron.  More empiric data is required.  However, we know that these techniques are indeed allowed in this one specific case gleaned from empiric knowledge of the universal molecular set in DNA.  In other words, if DNA symmetry were not broken precisely the way it is, the global symmetry relationships of actual codons would also be entirely different.

This is perhaps a good time to also recognize one more interesting geometric isomorphism in this particular scheme of illustration.  Recall that we constructed our dodecahedron first from a single tetrahedron.  However, the natural symmetry of this first tetrahedron allows for twelve distinct transformations or spatial rotations of the tetrahedron.  (Also note that each of these tetrahedrons has a mirror twin that is perhaps not relevant here.)

Figure 12.

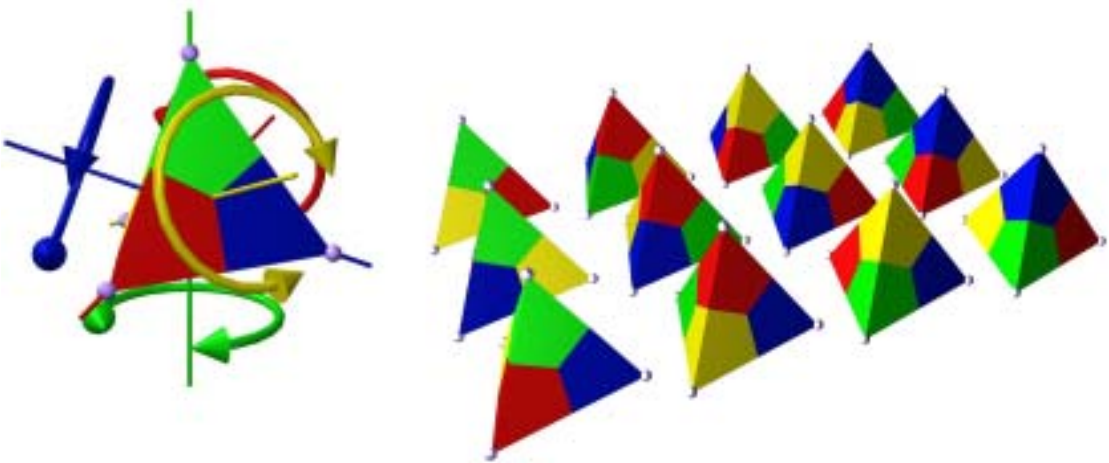

Furthermore, a tetrahedron when combined with its dual tetrahedron forms a cube. There are five interlocking cubes in a dodecahedron; therefore, there are 120 distinct transformations of a single tetrahedron within the points of a dodecahedron (2 X 5 X 12 = 120) not counting mirror twins. To help "see" this mathematical relationship we will need to add a fifth color to our initial illustration, the new color here being purple.

Figure 13.

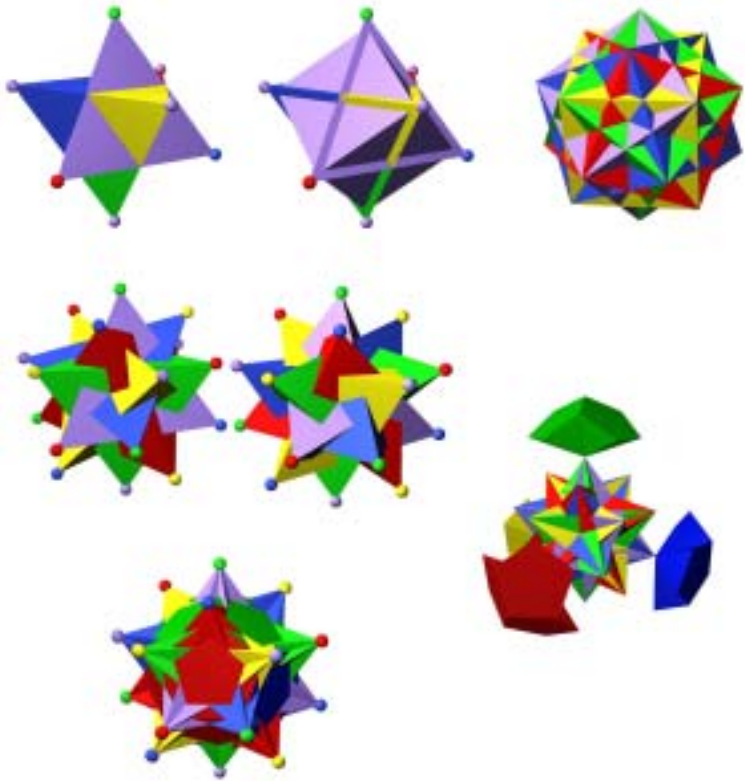

This proves that codon symmetry is not only isomorphic with all of the permutations of a dual triangle system; it is also isomorphic with all of the rotational permutations of a tetrahedron related to a single dodecahedron. In other words, this sequence symmetry can be perfectly extended into three-dimensional space.

Furthermore, combinations of any three contiguous faces of a dodecahedron can now precisely specify the spatial orientation of a single tetrahedron from the entire tetrahedral set.  In other words, the set of all triplet face sequences equal a second real set of structural orientations in this scheme.  All of the tetrahedrons are logically related to each other by consecutive permutations of the twelve dodecahedral faces.  So there also logically exist a minimum number of steps for getting from any one permutation to any other.  Something analogous to Hamiltonian circuits can represent specific relationships between dodecahedral faces to whole sets of tetrahedrons, so sequences are related to other sequences in a variety of logical ways.  Sequences can compete in terms of spatial efficiency.  Importantly, there now appears to be a primitive algebra of sequence and structure.  What these abstract observations mean is that a naturally occurring geometric language exists.  It translates dodecahedrons into sixty four unique tetrahedrons when using only permutations of four distinct face elements grouped symmetrically into threes.  This language exists independent of any set of objects that might somehow employ it in nature.  In other words, symmetry is the logical foundation of a sequence-structure language in this specific case.  This is the kind of logical foundation that nature could use to build a molecular information system that logically relates sequences to sequences, structures to structures, and structures to sequences.  There are an infinite number of ways it might specifically be done in nature.  There can be a fierce competition in finding "the best way."

It is interesting to note that the spatial symmetry of DNA's double helix can also easily be idealized as a sequence of dodecahedrons, and a protein is literally a sequence of amino acid tetrahedrons.  In other words, there is an undeniable spatial symmetry to the actual molecular components in the system used by nature, and it is isomorphic with its own sequence symmetry.  The fact that nature somehow found this natural geometric language as a basis for molecular sequence coding logic should, therefore, not be surprising to anyone.  For every codon in a set there can be a corresponding tetrahedron in a dodecahedron under the specific rules of this particular geometric information scheme.  Nature could easily use this as a natural basis of a molecular language in building molecules based purely on space-filling logic. It is perhaps a primitive glimpse into the central logic of a crystal computer. The basic rules of sequence, when employed for spatial information storage and translation, are entirely self-consistent. These rules were obviously in place here on earth before any molecules existed to provide us with the specific real world data that we like to study today as our metaphor of the genetic code.

The universe contains a primary logic that naturally relates sequence to structure. On the basis of symmetry alone, sequence and structure can become logically related. Sequence and structure can communicate information between molecular sets via common symmetry. There are many possible languages that can operate on this logic. More importantly, this should be the correct answer to the heretofore missing question of how structures in nature might make sequences and how structures can be informed by other sequences. Nucleotides are structures and proteins are structures, and they are mutually informed by the logical relationships between their own structures. Molecules must have languages and languages must have logic. It seems obvious that all molecules must at first be logically guided by their own structures. It seems even more obvious then that all molecular languages must at some level be languages of pure structure. After all, this is the only way for any molecule to "think" in general. This is the only way for any molecule to consistently perform any code at all. In this particular case, the genetic code is a structural language that has become capable of producing sequences only because of the consistency and symmetry of the molecular structures that operate the language. In other words, structural purity is the path to molecular sequence. Molecules typically eschew lines, but if the lines are really only manifestations of perfect structures, the molecules will comply. Simple structures can now be stored and translated into more complex structures by logical relationships between molecular sequences.

The key question in molecular biology must always at first boil down to the correct logical relationship between sequence and structure. This relationship is comically interpreted incorrectly and inverted in virtually every setting today. This simple misinterpretation of reality is without limit in its negative epistemic consequences. It would now perhaps behoove us to repeat the mantra "structure always logically subsumes sequence." This is true because the set of possible structures of complex molecules is always larger than the subset of possible sequences. Sequences can, therefore, never be the sole cause of structural effects. We must then always conceptually understand and define the genetic code as the natural functions that logically relate sets of molecular sequences understood to always be composed entirely of molecular structures. In other words, structure determines structure, and structure determines sequence – just as we should logically know it to be. The now unexpectedly difficult task of understanding the genetic code becomes one of understanding complex sets of molecular structures. After all, even "simple" sequences of molecules, like a codon, truly represent a molecular structure in nature. The codon is at first informed by its structure. Its sequence is merely a subset of this information. To be sure, structures can be simplified to the point where sequence becomes the dominant part of that structure, but sequence can never subsume structure, and molecular information can never become "linear" in this way. The symmetry principle has not been violated in nature; only in our model of nature.

Until now, we have said virtually nothing about the actual data that nature has given us to study in this translation system. The discussion has been abstract, only about basic symmetry and simple ways to represent it. We have merely created some linguistic and visualization tools based on fundamental codon symmetry in conjunction with the unique nature of DNA's natural two-bit set when placed within the elements of solid geometry. This exercise has generated a nifty data container with virtually no data in it, apart from the specific set of DNA nucleotides.

Figure 14.

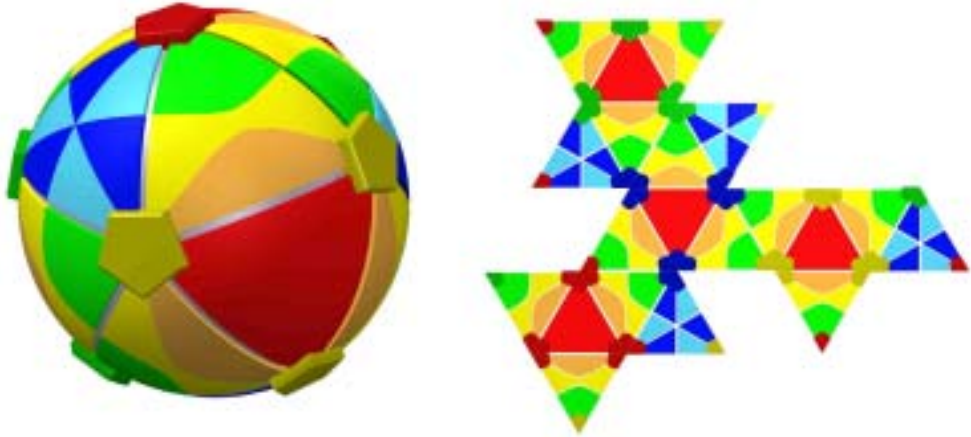

However, we can now see that the container itself forms obvious patterns based on the sequence symmetries that went into its construction. Not all codons are alike, but all codons do always inform each other. No codon ever has any meaning outside of the context of all other codons. Codons are a set that derives its meaning at first from its own structure and then from its logical organization relative to the structures of other molecular sets. From this treatment we can clearly see that codons can be logically spaced and inter-related based on symmetry. Individual permutations form logical subsets of permutations, and these subsets are inter-related; they too possess inherent symmetry. This particular codon system is particularly symmetrical in the sense that it efficiently packs codon symmetry into a coherent pattern within a context of DNA symmetry. We will now see that the actual data found in nature fits perfectly within those patterns.

**The Assignment of Amino Acids Conforms to the General Pattern of the Symmetry Group of actual Codons**

We will now use our new geometric visualization tools to analyze the real world data. I will illustrate and analyze the data within the context of these tools and argue that symmetry is the fundamental organizing principle behind the patterns we can see. I will present three forms of evidence to convince the reader that the data is in fact organized by the structures of codon symmetry. First, the evidence will be purely visual techniques based on properties of molecules within the data pattern. Second, the evidence will involve Gamow's remarkably perfect failure with respect to the predictions of his compact triangle model. Third, the evidence will be a handful of published findings that demonstrate diverse forms of symmetry that are acknowledged as valid forms of symmetry within this data set. Once the data has been illustrated and analyzed, I will argue that this treatment has tremendous epistemic value. The icon we choose can inform our thinking in productive ways.

As we examine the data that relates codons to amino acids we will need a property of amino acids to stand as "meaning" within the translation system. After all, apart from the context of an actual protein translation, an amino acid has no meaning in and of itself. Just as nucleotides derive their meaning from other nucleotides and codons derive their meaning from other codons, amino acids derive their meaning from relationships with other amino acids. Codons do not literally "mean" amino acids in the real world system of translation, but their assignment patterns demonstrate a remarkably consistent correlation across all known life forms. There is a broad, approximate symmetry of assignments between codons and amino acids for all life on earth. However, the basic problem of any codon map still holds here: amino acids are not the proper image of codons in translation. As long as amino acid sequences do not map to protein structures – and they clearly do not – then codons cannot mean only amino acids in translation. The symmetry principle cannot be violated for the mere sake of convenience here. In order to multiply the causes to at least match the effects, we will technically need to consider combinations of codons.  Unfortunately, this is well beyond the scope of this treatment, and far too complex for any simple mapping of codons.  Therefore, we must resign ourselves here to mapping codons to amino acids despite the fact that it cannot be a comprehensive map of translation.  It does, however, serve us well as a partial mapping of a demonstrably important subset of information translated by the genetic code.

For this analysis I have chosen to focus primarily on the property of amino acids known as water affinity.  This is only one dimension of amino acid meaning that surely must be symmetric with all others.  But for the time being, water affinity will stand as "meaning" within the translation system.  This is but one of many properties that we could have chosen, but it is a demonstrably important property when amino acids combine in sequences to form peptide bonds and then ultimately become proteins.  This property in the pattern of codon assignments can clearly be shown to reflect a tremendous amount of symmetry in the overall system of translation.  That is precisely what we will now do: we will find that symmetry in the assignments.

A quick look at the water affinities within the set of standard amino acids reveals that nature has selected a set that displays a smooth gradient with respect to water affinities across the entire set[14].  Amino acids are fairly well ordered with respect to water affinity.  I will use color once again as a tool to illustrate the data, but this time I will place water loving (hydrophilic) amino acids in the blue part of the color spectrum, and water hating (hydrophobic) amino acids in the red part of the color spectrum.  I have further used purple as a natural splice between the extreme water hating and extreme water loving amino acids to create a symmetrical color distribution, just like a color wheel.

Table 5.

| Water Affinity | | PK | | Name | Pos 1 | Pos 2 | Pos 3 | Missing |
|---|---|---|---|---|---|---|---|---|
| Highly Hydrophobic | 1 | 3.1 | | Isoleucine | A | U | ACU | G |
| | 2 | 2.5 | | Phenylalanine | U | U | CU | AG |
| | 3 | 2.3 | | Valine | G | U | ACGU | |
| | 4 | 2.2 | | Leucine | CU | U | ACGU | |
| | 5 | 1.1 | | Methionine | A | U | G | C |
| | 6 | 1.0 | | Tryptophan | U | G | G | AC |
| | 7 | 1.0 | | Alanine | G | C | ACGU | |
| | 8 | 0.67 | | Glycine | G | G | ACGU | |
| | 9 | 0.17 | | Cysteine | U | G | CU | A |
| | 10 | 0.08 | | Tyrosine | U | A | CU | G |
| | 11 | -0.29 | | Proline | C | C | ACGU | |
| | 12 | -0.75 | | Threonine | A | C | ACGU | |
| | 13 | -1.1 | | Serine | AU | CG | ACGU | |
| | 14 | -1.7 | | Histidine | C | A | CU | G |
| | 15 | -2.6 | | Glutamate | G | A | AG | C |
| | 16 | -2.7 | | Asparagine | A | A | CU | G |
| | 17 | -2.9 | | Glutamine | C | A | AG | U |
| | 18 | -3.0 | | Aspartate | G | A | CU | |
| Highly Hydrophilic | 19 | -4.6 | | Lysine | A | A | AG | CU |
| | 20 | -7.5 | | Arginine | AC | G | ACGU | |
| | | | | STOP | U | AG | AG | C |

Amino acid water affinity is a valid property to use for analysis here because it is such an important factor in the ultimate form that is taken by any protein structure. Furthermore, we will quickly see that it has played a key role in the symmetrical pattern of organization seen in the global codon assignments.

There are two factors to consider when we break the symmetry of any codon. Both factors play a significant role in the global assignment pattern for all codons with respect to amino acid water affinity. First, and most obvious, we must consider the identity of each nucleotide in the codon. Second, and less obvious, we must consider the position of each nucleotide identity within the order of each sequence. This means that there are actually twelve distinct nucleotide values within a specific context for every specific nucleotide sequence. The abstract principle is very familiar to us from our intuitive use of positional values in common numerical systems. So, one good way to visualize it is to look at the following set of twelve integers:

{1, 2, 3, 4, 10, 20, 30, 40, 100, 200, 300, 400}

It is not hard to find the pattern in this set, determine the rules that generated it, and recognize its obvious order. But it might be slightly harder to recognize what it tells us by analogy with nucleotides and their positional value within codons. In this case, four digits when combined with positional values – represented here by zeros included and omitted – will generate a set of twelve distinct integers. However, it is not hard for us to now imagine another set of sixty-four combinations of these integers that make a new set of sixty-four distinct integers. This new set would be recognizable as another ordered set based solely on symbol identity and position. So too can a set of codons be ordered in many different ways. We will explore but a few of them here.

The real meaning of nucleotide position in nature is perhaps not obvious until one considers the natural symmetry of any codon. Every codon must exist within the context of every other codon. All codons in nature are actual sequences of nucleotides. Actual sequences of nucleotides cannot avoid being transformed through time. The position of a nucleotide before a transformation will impact its set of possible states after transformation. The middle position is most prominent in assignment patterns because it anchors the symmetry of all codons before and after transformations. In other words, just as all codons are not equal because of their inherent symmetries, not all nucleotide positions are equal because of their inherent symmetries. This means that with respect to codon assignments, we can identify twelve distinct nucleotide values, one for each of the four types in each of the three positions. These two factors form two hierarchies with respect to water affinity within codon assignments - nucleotide identity and position:

Nucleotide Indetity

1. A – Adenine (1)
2. C – Cytosine (4)
3. G – Guanine (9)
4. U – Uracil (16)

Nucleotide Position

1. 2$^{nd}$ Position (3)
2. 1$^{st}$ Position (2)
3. 3$^{rd}$ Position (1)

Using these two hierarchies and the somewhat arbitrary weighting values given here we can demonstrate that the assignments within the set of codons reflect a complex yet obvious pattern of amino acid water affinities. In other words, the amino acids can be ordered, the nucleotides can be ordered, the codons can be ordered, and the ordering of all three sets can be related to each other. It may surprise some to learn that similar techniques must also lie behind any codon table. Any table is at bottom a kind of mathematical formula to weight and thereby linearly arrange codons. We just fail to recognize the formal method of ordering within the standard spreadsheet of codons, probably because nobody has ever perceived any real use for it. To be honest, the one found most often in print is not a very good one, and we can easily improve upon it. We can produce an alternate arrangement within that same structure by merely substituting these new weighting values into the variables for the same codon weighting formula that is covertly used to arrange the standard codon table in most textbooks. Of course, we will also then lose the clever partial compression of nucleotide symbols that makes the table so convenient in the first place.

Figure 15.

$$\text{Codon} = \sum_{x=1}^{x=3} (\text{Position Value}(x) \times \text{Nucleotide Value}(x))$$

Nucleotide Values

A [1]  C [4]  G [9]  U [16]

Position Values

#1 [2]  #2 [3]  #3 [1]

| 96 UUU Phe | 89 UUG Leu | 84 UUC Phe | 82 GUU Val |
| 81 UUA Leu | 75 UGU Cys | 75 GUG Val | 72 CUU Leu |
| 70 GUC Val | 68 UGG Trp | 67 GUA Val | 66 AUU Ile |
| 65 CUG Leu | 63 UGC Cys | 61 GGU Gly | 60 CUC Leu |
| 60 UCU Ser | 60 UGA STP | 59 AUG Met | 57 CUA Leu |
| 54 AUC Ile | 54 GGG Gly | 53 UCG Ser | 51 UAU Tyr |
| 51 CGU Arg | 51 AUA Ile | 49 GGC Gly | 48 UCC Ser |
| 46 GCU Ala | 46 GGA Gly | 45 AGU Ser | 45 UCA Ser |
| 44 UAG STP | 44 CGG Arg | 39 UAC Tyr | 39 CGC Arg |
| 39 GCG Ala | 38 AGG Arg | 37 GAU Asp | 36 CCU Pro |
| 36 UAA STP | 36 CGA Arg | 34 GCC Ala | 33 AGC Ser |
| 31 GCA Ala | 30 ACU Thr | 30 AGA Arg | 30 GAG Glu |
| 29 CCG Pro | 27 CAU His | 25 GAC Asp | 24 CCC Pro |
| 23 ACG Thr | 22 GAA Glu | 21 AAU Asn | 21 CCA Pro |
| 20 CAG Gln | 18 ACC Thr | 15 CAC His | 15 ACA Thr |
| 14 AAG Lys | 12 CAA Gln | 9 AAC Asn | 6 AAA Lys |

We have merely achieved a new arrangement for the standard presentation of data. However, we can now change the arrangement scheme slightly and present the data in a pseudo "linear" format. In so doing, we can see that the smooth gradient, or rainbow of water affinities has in some way been preserved in the assignment of amino acids across the entire codon set. More importantly, we can begin to guess that some arrangements of this data can be more useful to us than are others.

Figure 16.

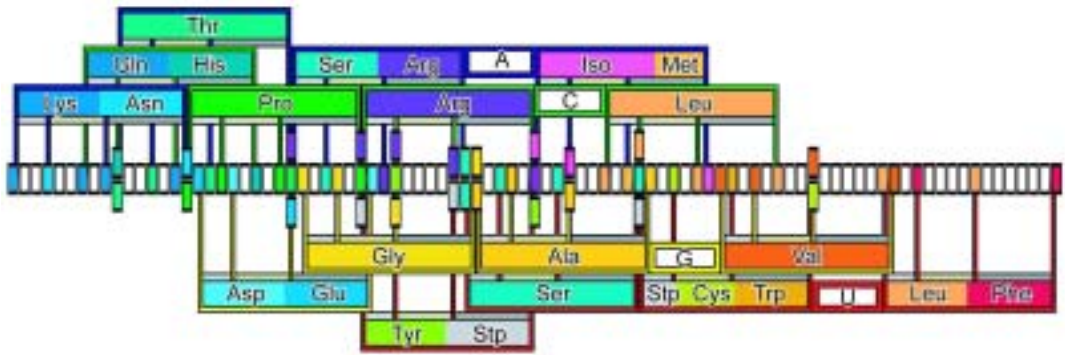

We see that this is not merely a single rainbow but somehow a weaving of many rainbows based on the four multiplets that make up the four nucleotide poles of the codon group. This is a complex rainbow indeed, but I do not feel that it is best illustrated within the context of any standard linear table. The water affinities of amino acids do indeed form a matrix of interrelated assignments not a single line of assignments. The codon table is merely a single slice of a more complex pattern, yet the standard method of arrangement is entirely subjective and asymmetrical. It represents an inherently limited approach toward illustrating the data and its global symmetry.

The actual assignment pattern in nature is, therefore, better viewed as a symmetrical matrix of assignment patterns. If a table of this sort is to be used, then multiple tables should be produced, one for each symmetry. However, to fully appreciate the natural beauty of this arrangement we can use the illustration tools of symmetry that we created earlier. A single pattern with symmetry is better here than many patterns without. Codon assignments are primarily a relationship built upon a complex symmetry within the data, and these tools illustrate the symmetry as well as the data. We will start with the categorization of codon classes and types, and illustrate the distribution of amino acids based on those categories alone.

Figure 17.

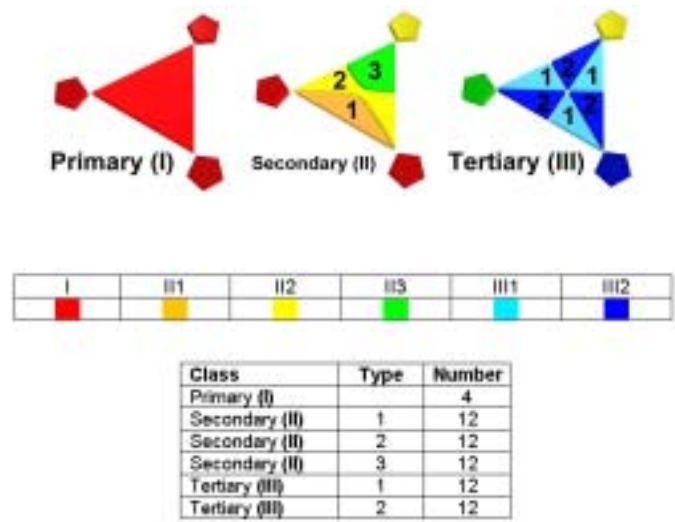

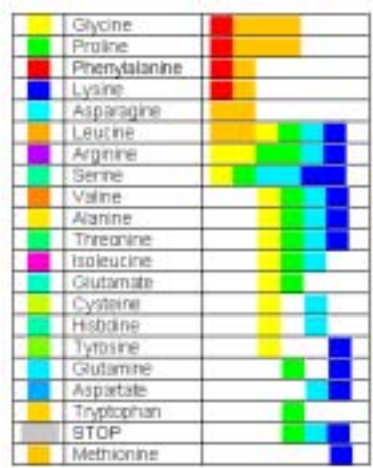

This too fails to fully illuminate the complex pattern of water affinities as they relate to codon types. A much better way to illustrate the global pattern based on codon class and type is to develop an entirely different weighting scheme, one that better respects the contribution of every nucleotide in every position. In other words, we need a scheme that reflects the fact that there are actually twelve different individual nucleotide values in the set of real codons. I have chosen to use continued fractions to create this new codon weighting scheme with integer values.

Figure 18.

$$\text{Codon} = B2 + \cfrac{1}{B1 + \cfrac{1}{B3}}$$

We can now generate a rational fraction, or a numerator and denominator for every codon, organize each codon within its class based on its weight, and then easily see that each codon class and type forms a credible rainbow with respect to amino acid assignments and their water affinity. In other words, the complex symmetry of the codon set has captured a complex symmetry of amino acid water affinities in making global codon assignments. This observation is completely lost in the standard table.

Figure 19.

**Primary (I)**

| CF | Icon | Codon | N | D | Dec | AA | Water |
|---|---|---|---|---|---|---|---|
| 1;1,1 | | AAA | 3 | 2 | 1.5 | Lys | |
| 2;2,2 | | CCC | 12 | 5 | 2.4 | Pro | |
| 3;3,3 | | GGG | 33 | 10 | 3.333 | Gly | |
| 4;4,4 | | UUU | 72 | 17 | 4.235 | Phe | |

**Secondary Type 1 (II1)**

| CF | Icon | Codon | N | D | Dec | AA | Water |
|---|---|---|---|---|---|---|---|
| 1;1,2 | | AAC | 5 | 3 | 1.666 | Asn | |
| 1;1,3 | | AAG | 7 | 4 | 1.75 | Lys | |
| 1;1,4 | | AAU | 9 | 5 | 1.8 | Asn | |
| 2;2,1 | | CCA | 7 | 3 | 2.333 | Pro | |
| 2;2,3 | | CCG | 17 | 7 | 2.429 | Pro | |
| 2;2,4 | | CCU | 22 | 9 | 2.444 | Pro | |
| 3;3,1 | | GGA | 13 | 4 | 3.25 | Gly | |
| 3;3,2 | | GGC | 23 | 7 | 3.286 | Gly | |
| 3;3,4 | | GGU | 43 | 13 | 3.308 | Gly | |
| 4;4,1 | | UUA | 21 | 5 | 4.2 | Leu | |
| 4;4,2 | | UUC | 38 | 9 | 4.222 | Phe | |
| 4;4,3 | | UUG | 55 | 13 | 4.231 | Leu | |

**Secondary Type 2 (II2)**

| CF | Icon | Codon | N | D | Dec | AA | Water |
|---|---|---|---|---|---|---|---|
| 1;4,4 | | UAU | 21 | 17 | 1.235 | Tyr | |
| 1;3,3 | | GAG | 13 | 10 | 1.3 | Glu | |
| 1;2,2 | | CAC | 7 | 5 | 1.4 | His | |
| 2;4,4 | | UCU | 38 | 17 | 2.235 | Ser | |
| 2;3,3 | | GCG | 23 | 10 | 2.3 | Ala | |
| 2;1,1 | | ACA | 5 | 2 | 2.5 | Thr | |
| 3;4,4 | | UGU | 55 | 17 | 3.235 | Cys | |
| 3;2,2 | | CGC | 17 | 5 | 3.4 | Arg | |
| 3;1,1 | | AGA | 7 | 2 | 3.5 | Arg | |
| 4;3,3 | | GUG | 43 | 10 | 4.333 | Val | |
| 4;2,2 | | CUC | 22 | 5 | 4.4 | Leu | |
| 4;1,1 | | AUA | 9 | 2 | 4.5 | Iso | |

## Secondary Type 3 (II3)

| CF    | Icon | Codon | N  | D  | Dec   | AA   | Water |
|-------|------|-------|----|----|-------|------|-------|
| 1;4,1 |      | UAA   | 6  | 5  | 1.2   | STOP |       |
| 1;3,1 |      | GAA   | 5  | 4  | 1.25  | Glu  |       |
| 1;2,1 |      | CAA   | 4  | 3  | 1.333 | Gln  |       |
| 2;4,2 |      | UCC   | 20 | 9  | 2.222 | Ser  |       |
| 2;3,2 |      | GCC   | 16 | 7  | 2.286 | Ala  |       |
| 2;1,2 |      | ACC   | 8  | 3  | 2.666 | Thr  |       |
| 3;4,3 |      | UGG   | 42 | 13 | 3.231 | Trp  |       |
| 3;2,3 |      | CGG   | 24 | 7  | 3.429 | Arg  |       |
| 3;1,3 |      | AGG   | 15 | 4  | 3.75  | Arg  |       |
| 4;3,4 |      | GUU   | 56 | 13 | 4.308 | Val  |       |
| 4;2,4 |      | CUU   | 40 | 9  | 4.444 | Leu  |       |
| 4;1,4 |      | AUU   | 24 | 5  | 4.8   | Iso  |       |

## Tertiary Type 1 (III1)

| CF    | Icon | Codon | N  | D  | Dec   | AA   | Water |
|-------|------|-------|----|----|-------|------|-------|
| 1;4,3 |      | UAG   | 16 | 13 | 1.231 | STOP |       |
| 1;3,2 |      | GAC   | 9  | 7  | 1.286 | Asp  |       |
| 1;2,4 |      | CAU   | 13 | 9  | 1.444 | His  |       |
| 2;4,1 |      | UCA   | 11 | 5  | 2.2   | Ser  |       |
| 2;3,4 |      | GCU   | 30 | 13 | 2.308 | Ala  |       |
| 2;1,3 |      | ACG   | 11 | 4  | 2.75  | Thr  |       |
| 3;4,2 |      | UGC   | 29 | 9  | 3.222 | Cys  |       |
| 3;2,1 |      | CGA   | 10 | 3  | 3.333 | Arg  |       |
| 3;1,4 |      | AGU   | 19 | 5  | 3.8   | Ser  |       |
| 4;3,1 |      | GUA   | 17 | 4  | 4.25  | Val  |       |
| 4;2,3 |      | CUG   | 31 | 7  | 4.429 | Leu  |       |
| 4;1,2 |      | AUC   | 14 | 3  | 4.666 | Iso  |       |

## Tertiary Type 2 (III2)

| CF    | Icon | Codon | N  | D  | Dec   | AA   | Water |
|-------|------|-------|----|----|-------|------|-------|
| 1;4,2 |      | UAC   | 11 | 9  | 1.222 | Tyr  |       |
| 1;3,4 |      | GAU   | 17 | 13 | 1.308 | Asp  |       |
| 1;2,3 |      | CAG   | 10 | 7  | 1.429 | Gln  |       |
| 2;4,3 |      | UCG   | 29 | 13 | 2.231 | Ser  |       |
| 2;3,1 |      | GCA   | 9  | 4  | 2.25  | Ala  |       |
| 2;1,4 |      | ACU   | 14 | 5  | 2.8   | Thr  |       |
| 3;4,1 |      | UGA   | 16 | 5  | 3.2   | STOP |       |
| 3;2,4 |      | CGU   | 31 | 9  | 3.444 | Arg  |       |
| 3;1,2 |      | AGC   | 11 | 3  | 3.666 | Ser  |       |
| 4;3,2 |      | GUC   | 30 | 7  | 4.286 | Val  |       |
| 4;2,1 |      | CUA   | 13 | 3  | 4.333 | Leu  |       |
| 4;1,3 |      | AUG   | 19 | 4  | 4.75  | Met  |       |

Of course, the codon classes and types are merely a property of the symmetry group of codons as illustrated above. It then seems valid from this alone to conclude that the assignment of amino acids is somehow predicated on the symmetry of codons. However, to fully appreciate this global symmetry and what it means, or how it has been deployed within the much larger translation system by nature, we will now need to rely heavily on the global visualization techniques developed above. After all, it is a complex task to dissect out codon symmetries when those symmetries are based on actual transformations of whole codon sequences. Indeed, it is such a complex system of symmetry relationships that we can only hope to visualize it by dissecting out components within the larger context of perfect symmetry. We will start by filling our generic data container – the G-ball - with the empiric assignments of amino acids.

Figure 20.

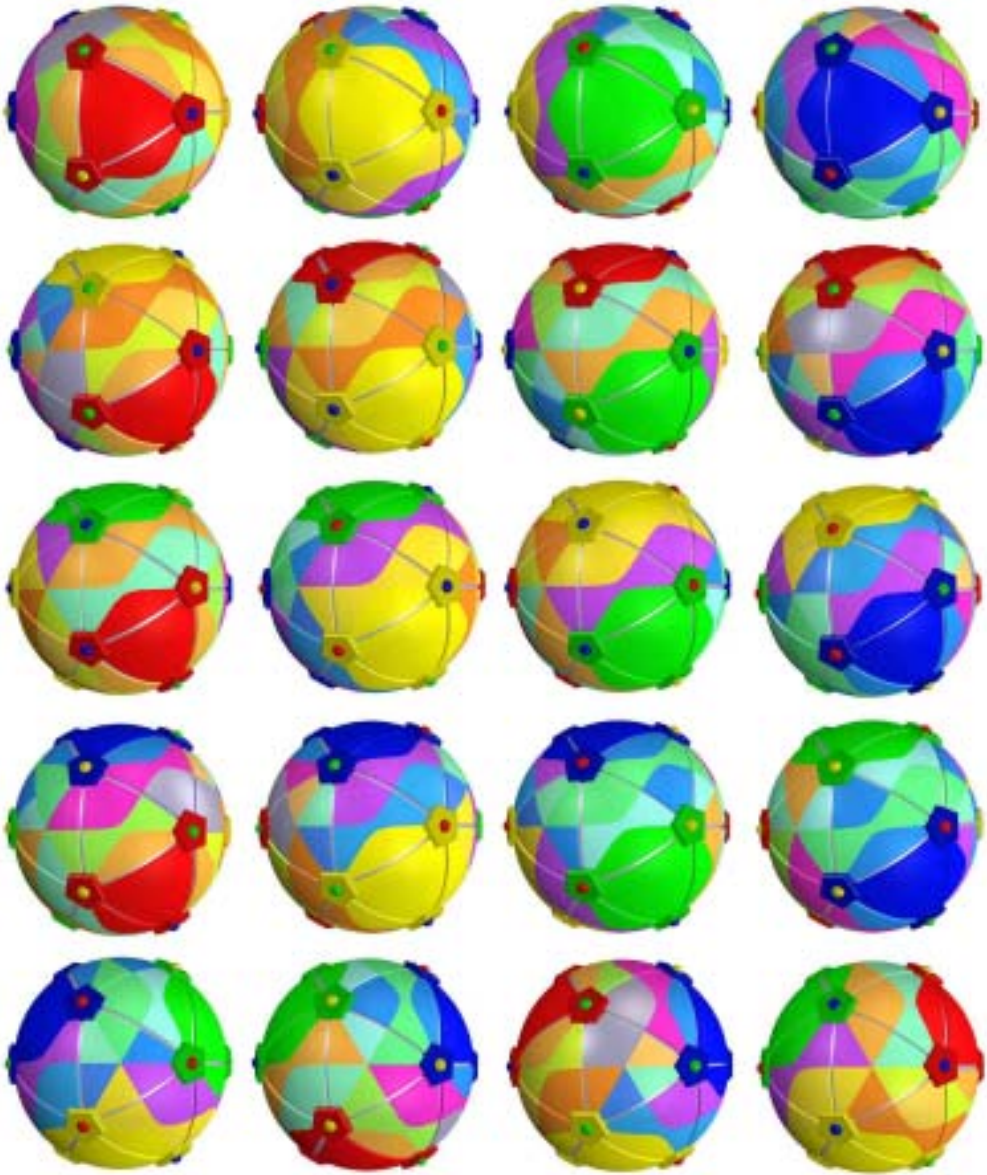

Unlike a table or a line, the G-ball is an un-weighted arrangement of this set of codons. There is almost no subjectivity to the placement of any component of the system relative to the other components in the system. There are only two ways to place the twelve nucleotides, this way and its mirror that swaps any two sets of three similar nucleotides. The rest of the components must fall where they fall. Although some codons appear to be treated differently than others, they are not. All codons and all nucleotides are treated exactly the same. The amount of space on the map occupied by each codon and the relative positions between codons are merely measures of their inherent symmetry.

This does not perhaps reveal as compelling of a rainbow pattern because it actually reflects an interwoven matrix of many different patterns. So, we will begin dissecting them out individually by first tracing the obvious rainbow of water affinities that we now know exists within the codon group. We know it is there because we have just seen it, but where is it on this particular map? We will start with the four major nucleotide poles and their resulting multiplets to "fold the rainbow" into its requisite tetrahedron. This appears to require a smattering of oddball assignments to seemingly act as glue within the global pattern.

Figure 21.

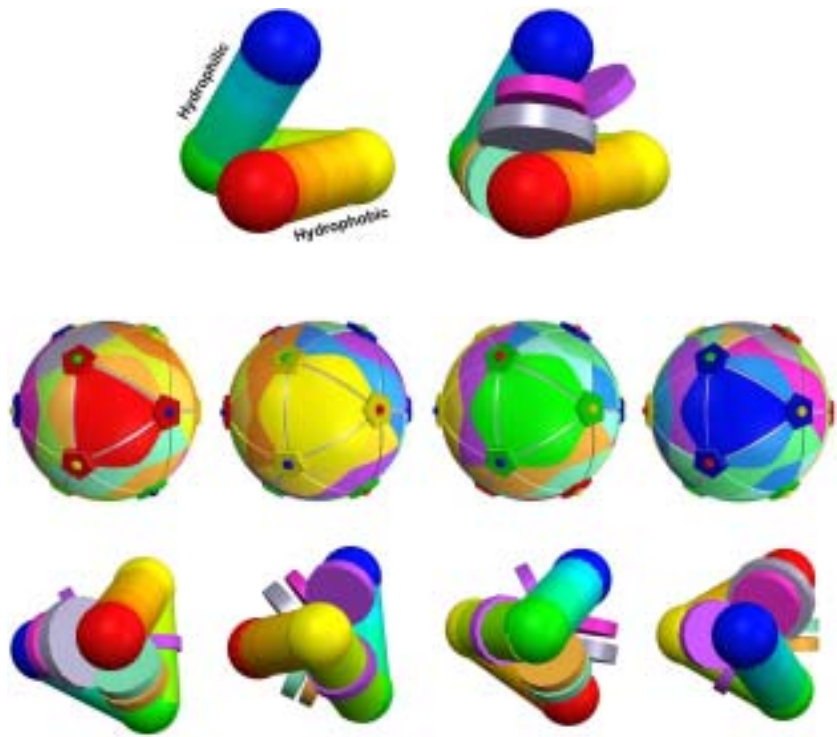

This is far from perfect, merely a gross visual tracking device to identify the complex general rainbow within the globally symmetric data container. The rainbow remains somewhat hidden in this form, but we know it is there and we can still see large parts of it. However, we can now see that the rainbow has a beginning, middle and end, and the beginning has been folded by nature back to meet with the end within this symmetric container. Codons form an ordered set with respect to their assignments to the rainbow of amino acid water affinity. It acts just like a musical scale or just like a color wheel in joining beginnings with endings of a single spectrum. This makes more sense when we begin to view codons and their assignments as a form of standing wave[15]. It is a dynamic process of sequence generation when sequences are taken within the larger context of all sequences over all time. Nodes of stability will form within the larger pattern. We can more easily see this complex relationship when we unfold the multiplets and display them in a simple series of the four major poles.

Figure 22.

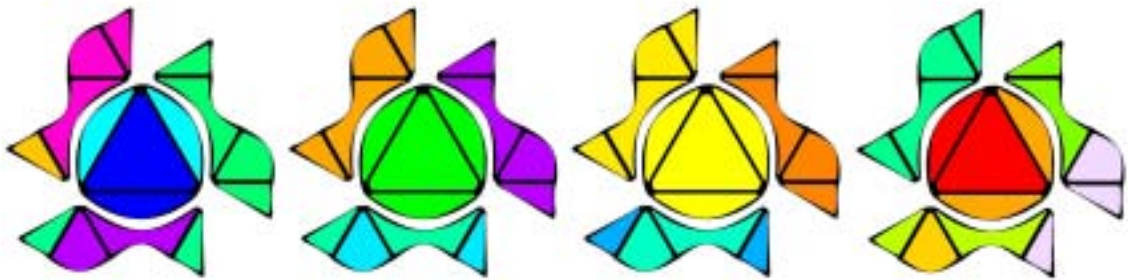

From this we can see the contribution of each multiplet to the overall rainbow pattern. We can see that the start codon on the far left initiates the pattern that then generally proceeds from water loving to water hating and terminates in a tight pattern of the three stop codons. The start and stop codons form a "wall" or perhaps a "splice" between hydrophobic and hydrophilic codons in the folded series. The whole pattern can be seen as a complex rainbow continuum with a beginning, middle and end, and the beginning and end merely wrap around the color wheel to join the two extremes of the pattern. In this context, the genetic code has apparently used start and stop codons to splice the continuum in the same way that nature uses purple to splice red and blue on opposite extremes of the color spectrum into a perfect circle.

However, we must be ever-mindful that codons do not mean amino acids, and water affinity is but one property in a more complex overall scheme. Notably, codons for proline and glycine clearly form the strongest sub-pattern in the overall pattern, and they are perfectly balanced within their dominant assignments in the very middle of the continuum. Proline and glycine are each assigned an entire homogeneous multiplet that is perfectly symmetrical with the other. But remember, this map is but one simple representation of symmetry within a vastly more complex manifestation of symmetry. However, because of the known symmetry of DNA, the strongest patch of symmetry within this set of codons resides between the G and C poles of all codons. Besides water affinity, proline and glycine provide a strong duality of meaning as it relates to the structural properties of amino acids in general, especially when they combine in sequence to form protein structures, like loops and turns. These two amino acids represent a complementary "swivel" and "latch" motif in the polypeptide backbone, and they can be symmetrically positioned in a sequence to do this. In other words it does not matter so much that proline comes before glycine in a sequence, just that they appear together. This particular arrangement of amino acid assignments ensures that this configuration will occur in nature with the greatest consistency despite all sequence transformations. Conversely, the A:U pole is perfectly symmetric with respect to the extremes of water affinity. These are valuable complex symmetries that life can utilize during inevitable transformations of coding sequences, transformations that occur with certainty in DNA's replication and recombination. We can now plainly see this defining global symmetry to the assignments of the code itself. But we can only imagine how they are used in decoding these sequence symmetries if we view them from the context of a globally symmetric structure of the overall codon assignment pattern.

Perhaps a more convincing demonstration of the global symmetry pattern can be found in the individual symmetry transformations of the data itself. In other words, we can ask what will happen to entire codon sequences when all individual codons undergo the same transformations. We will start by looking at the logical impact each of these transformations has on the entire pattern of the generic data container. We will then show that the empiric data actually conforms in a remarkably consistent way to the many and varied patterns of these sequence transformations. We will use the multiplet arrangement of the major poles to visualize the impact of entire sequence transformations on the global assignment of codons. As a standard convention here we will put the C-pole in the center of the pattern.

Figure 23.

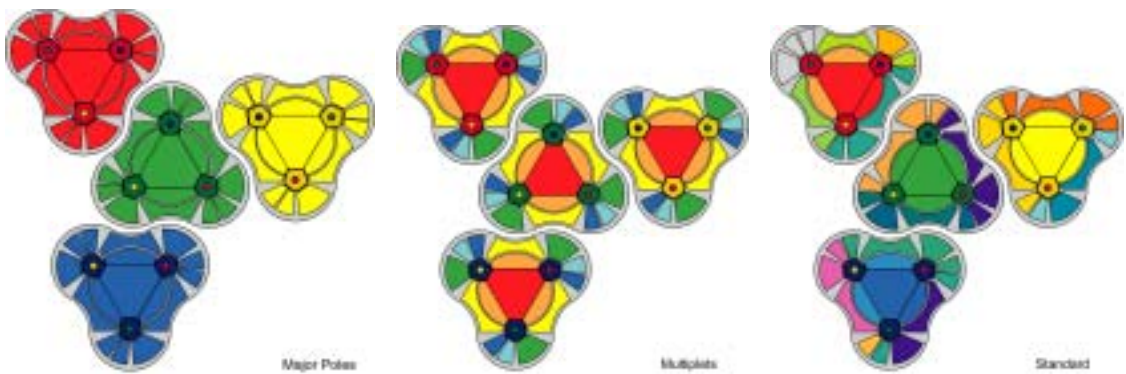

A genome is a sequence of individual nucleotides. It is transformed by sequence symmetry when new genomes are inevitably formed. Codon reading frames are shifted in both directions, they are complemented, they are inverted, point mutated, and combinations of all transformations are typically executed through time in nature. The genetic code is structured upon a global symmetry that is able to anticipate all of these transformations, which makes it a remarkably effective tool for consistently decoding genomes through time, genomes that must always result from all sequence transformations. This is the primary benefit to nature having organized the genetic code - and with it the assignment of amino acids – entirely around symmetry. We can use our visual tools of symmetry to see in part how this has actually been done. We will start by visualizing a short sequence of 101 random nucleotides.

Figure 24.

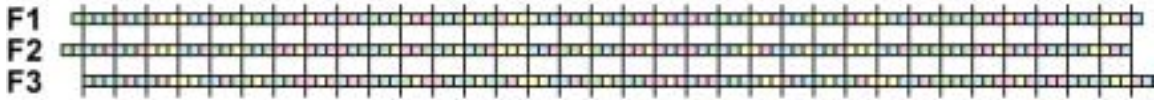

First codon in sequence

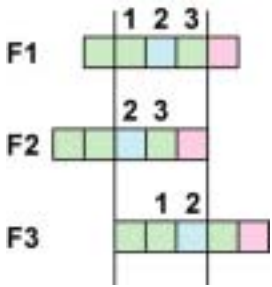

The first reading frame, F1, is the reference frame of identity symmetry. The second frame, F2, is shifted forward one nucleotide. This corresponds to the symmetry of one rotation of every codon in the sequence. The third frame, F3, is shifted backward one nucleotide, which also corresponds to symmetry of two rotations of every codon in the sequence. Note that all of the codons in the random sequence of 101 nucleotides are transformed in the same way during every transformation. However, since the sequence is random, there is no way to anticipate exactly which nucleotide will be removed and added to each codon after a frame shift occurs. It seems that it logically should be a randomizing event over the entire sequence, but we will see that the genetic code has taken advantage of codon symmetry to insure that sequence transformations maintain elements of protein information after transformations of all kinds. The code never "sees" individual codons. It sees entire codon sequences and entire sequence contexts. Only a globally symmetric assignment pattern can do this, and only a perfectly symmetric assignment pattern could account for all possible transformations simultaneously. This allows us to glimpse the value and meaning of symmetry within the global pattern of codon assignments. Codons derive their meaning from context, and the meaning of all codons is derived from all possible codon contexts. Symmetry is the foundation of all codon contexts.

We can start by examining the literal impact of a transformation on the two different kinds of multiplets within each major pole. We will use CCN (N stands for an unknown nucleotide) as the homogenous multiplet, and CGN as the heterogeneous multiplet for this illustration.

Figure 25.

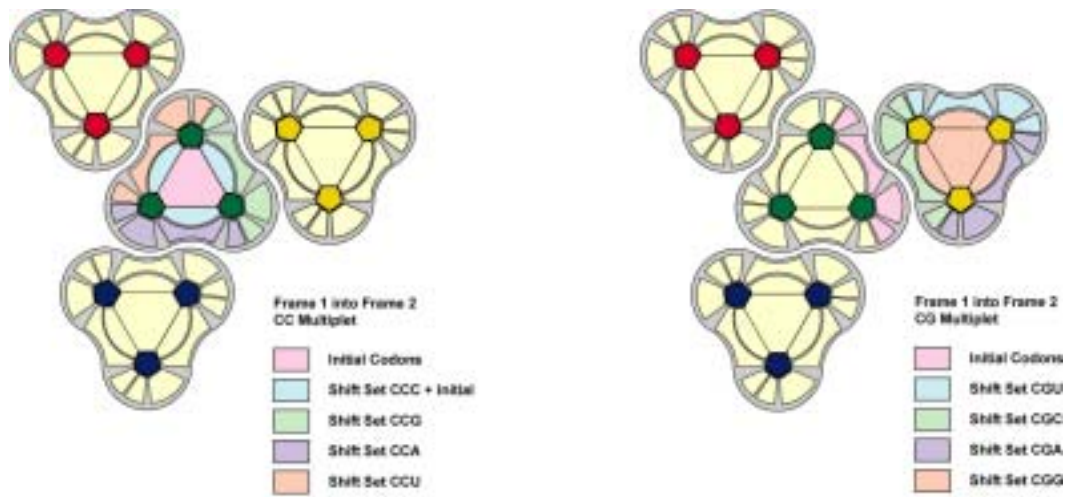

We can see that the scatter pattern of a forward shift in a homogeneous multiplet keeps the new codon entirely within the original pole. In other words, all CCN codons stay in the C-pole when they shift forward via r-symmetry. Type I codons all stay within the homogeneous multiplet. Each of the three type II1 codons creates four possible new codons, and this new group of four is tightly contained in adjacent heterogeneous multiplet of that pole. Conversely, the CGN heterogeneous multiplet begins a transformation with four different types of codons that each generate four entirely new codons when shifted. The group of four codons for each original CGN codon will fall into one of the four multiplets in the G-pole because G is the middle nucleotide in the original codon. All of this is confusing in words but should be apparent in the pictures.

The r-symmetry of codons is the same as common wobble symmetry easily recognized in nucleotide sequences. This type of symmetry has been apparent since the first day the codon table became known. One cannot help but see it because its pattern is so strong across the spectrum of assignments. However, this symmetry is frequently over-idealized and misinterpreted. The wobble groups are prominent in the genetic code because amino acid assignments are made with respect to r-symmetry, no doubt, but r-symmetry is not the only thing that dictates this global assignment pattern. There is vastly more symmetry within the assignments and it too plays an important role. The picture gets even more interesting when we examine the pattern of these other symmetries in the overall group, especially $r^2$-symmetry, which we will now visit.

Figure 26.

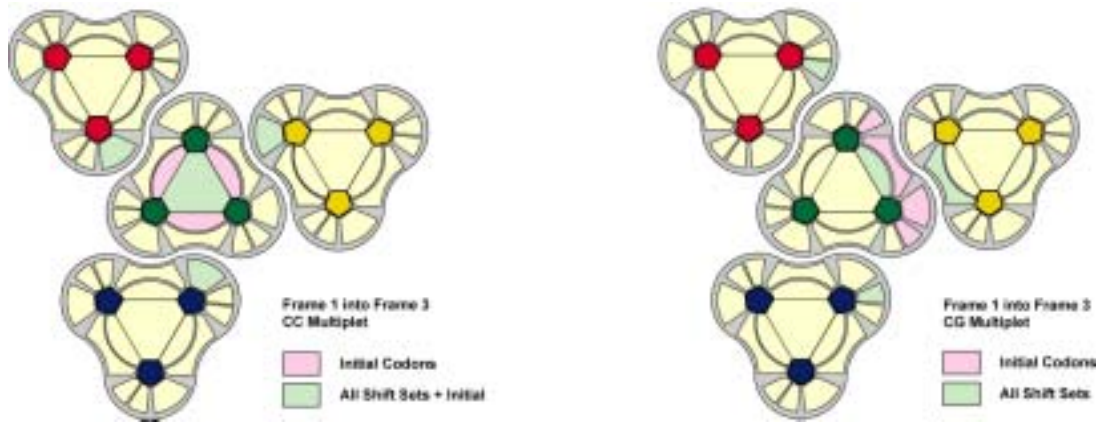

The scatter pattern of $r^2$-symmetry or a backward shift on the CCN multiplet stays within the type I codon and the type II3 codons from each of the other three poles. In other words, all four original codons shift into the same four shifted codons (NCC). Equally interesting is that each of the four CGN multiplets will shift backward into four different types of codon, II1, II2, III1 and III2, from four different poles, but it is the same four codons (NCG) for each of the original codons in the CGN multiplet. Whereas a multiplet shifts forward into sixteen codons, it shifts backward into just four. Unfortunately, this scatter pattern is still too complex globally to immediately appreciate the impact it has had on the global assignment pattern - like we so easily can with wobble. Fortunately, we can now use a nifty graphical trick to clarify this obvious impact. Note that each triplet has three nucleotides and one of those nucleotides is removed from the triplet in a shift transformation. But that same nucleotide is one of the four possible replacements in the triplet after the shift. It is, after all, a cyclic permutation. We can, therefore, merely replace that shifted nucleotide in the permutation for the new codon, and we can do this for every codon on the map. For instance, CGA becomes ACG. When we do this for $r^2$-symmetry of every codon, the overall pattern of the map now looks like this:

Figure 27.

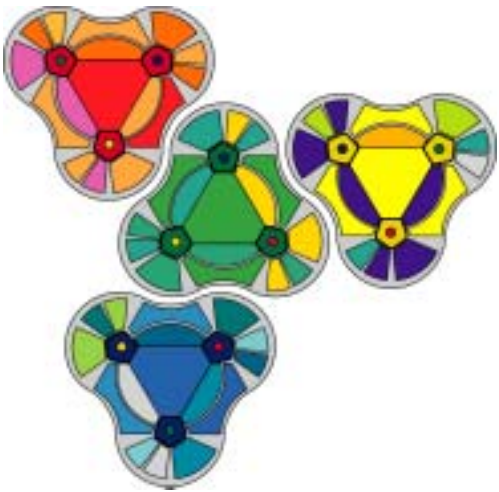

And the rainbow series of major poles after global $r^2$ codon replacement looks like this:

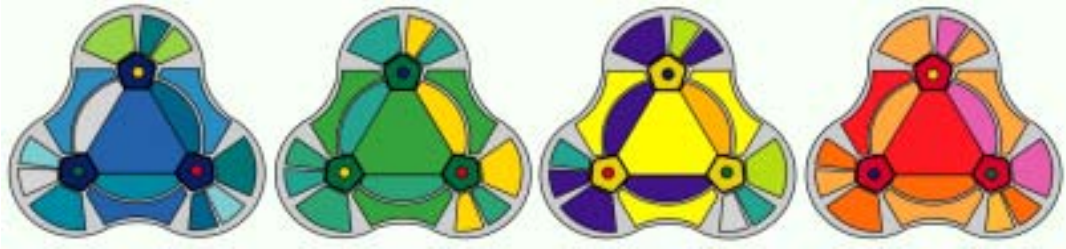

This should give one pause. It is a stunningly consistent pattern with respect to global amino acid assignments, major nucleotide poles, water affinity and $r^2$-symmetry of every single codon. In other words, $r^2$-symmetry has apparently played an identifiable role in making global amino acid assigments across all codons. Backward and forward shifts now cooperate in the global assignment pattern. I don't know what constitutes proof of this, but this graph is all the proof I need to conclude that these particular assignments were made based in part on $r^2$-symmetry. Although accounting for the universally acknowledged r-symmetry of codons is a simple matter of respecting multiplets in the assignment pattern, accounting for $r^2$-symmetry is a far more complex matter of weaving together all sixteen of the multiplets. This diagram reflects the fact that this has, in fact, actually been done in nature.

However, there are still other symmetries to consider in the pattern. Take, for instance, reflection symmetry of a sequence, or $mr^2$-symmetry. When we double rotate and then reflect a sequence of three nucleotides we merely create the inverse order of the original codon. In other words, 1-2-3 becomes 3-2-1 and there need be no new nucleotides in the codon, so to see the impact of reflection symmetry on the global pattern we merely need to replace every codon on the map with its order inverse.

Figure 28.

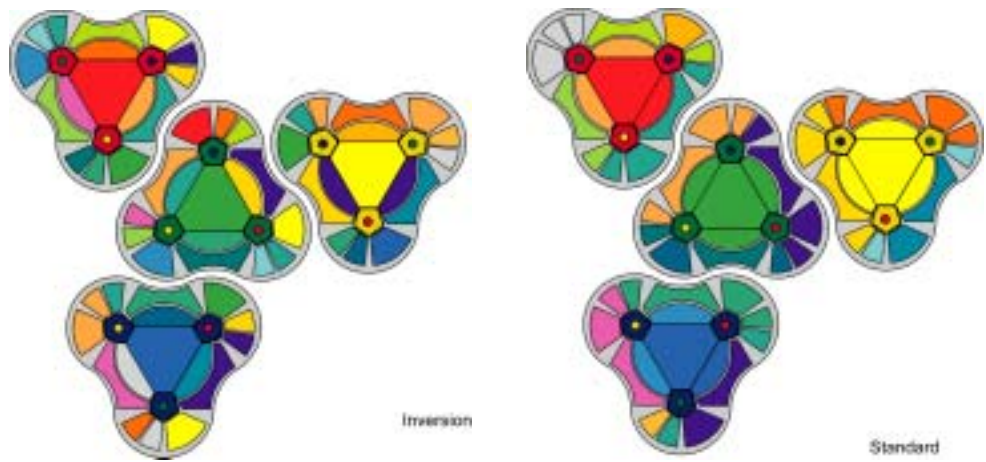

Every codon shares a Cayley triangle with its inverse because of reflection sequence symmetry. Replacing codons with their order inverse, therefore, is an exercise in shuffling and rotating every triangle. This should seem to somehow randomize and greatly fragment the overall pattern, but as we can see from the above diagrams, the pattern remains remarkably consistent after the transformation is globally performed. This is due in part to the fact that some permutations merely rotate into themselves, but also it is due to a global symmetry of assignments. One might naturally speculate that this is probably extremely useful in the real world of molecules where genomes frequently become palindromes.

Still more remarkable, perhaps, is the fact that this codon system has become a dual-binary system merely because it incorporates two complement pairs of nucleotides. This means that there is another type of reflection symmetry within the system – or inverse symmetry: the reflected symmetry of complement pairing. The reflected symmetry of complements is perhaps a more impressive example of global symmetry because it is embedded throughout the fundamental structure of the entire system. We are not transforming codons into their image – anticodons – but we are transforming them into their complementary sequences on the non-coding strand. This happens in the real world with high frequency of recombining genomes. Non-coding strands frequently become coding strands because they already exist. We can see this impact on assignments by performing a similar graphic trick with the entire map.

Figure 29.

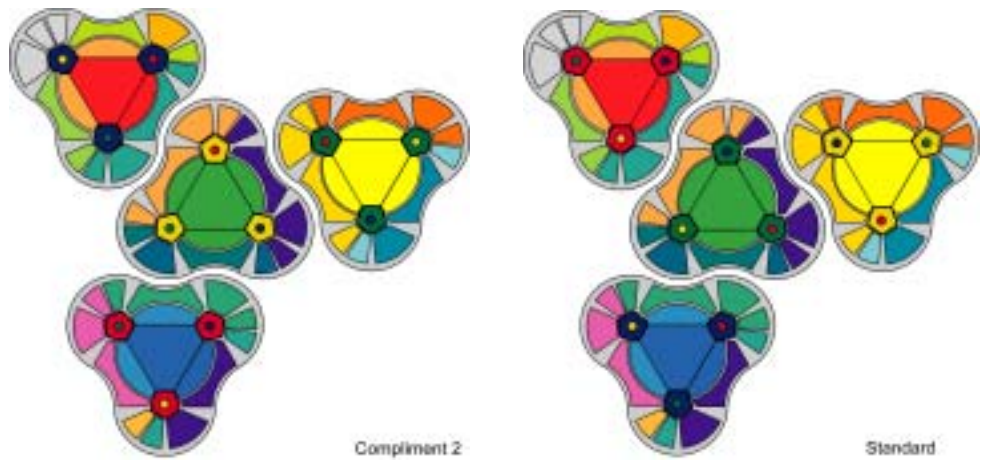

These assignment patterns are, of course, identical. The graphical trick performed here is merely to replace every codon with its complement and then rearrange the four major poles, swapping C with G, and A with U. The complement symmetry of codons is reflected in the complement symmetry of DNA. It is simply a property of the overall system. This trick is only possible with this specific set of nucleotides, or a dual binary system of information. However, note that the amino acids in the A pole are generally complementary with the amino acids in the U pole, and those in the C pole are complementary with the G pole. The reflected symmetry of complementary codons is represented in the properties of the amino acids to which they are assigned. Once again, it is an incredibly symmetric assignment pattern when all codons are considered globally.

Some complain that we are merely "playing with the data." However, this is nature's data, and only within the context of global symmetry can we make this data seem to sit up and deal cards, so to speak. These tricks are tricks of nature not of data manipulation per se. In fact, more play needs to occur simply because so much more play is possible within this context. It is a system built for play because it is a competitive system. But besides these specific sequence symmetry transformations, there are other ways to detect a global symmetry in the codon assignments. Consider the case of point mutations. We can use them to examine the effects of partial randomization in the global pattern. For instance, a point mutation involves the random change of any nucleotide in any position in any codon. We can see the randomizing effects of all point mutations when applied to one homogeneous and one heterogeneous multiplet from the C-pole of the data container.

Figure 30.

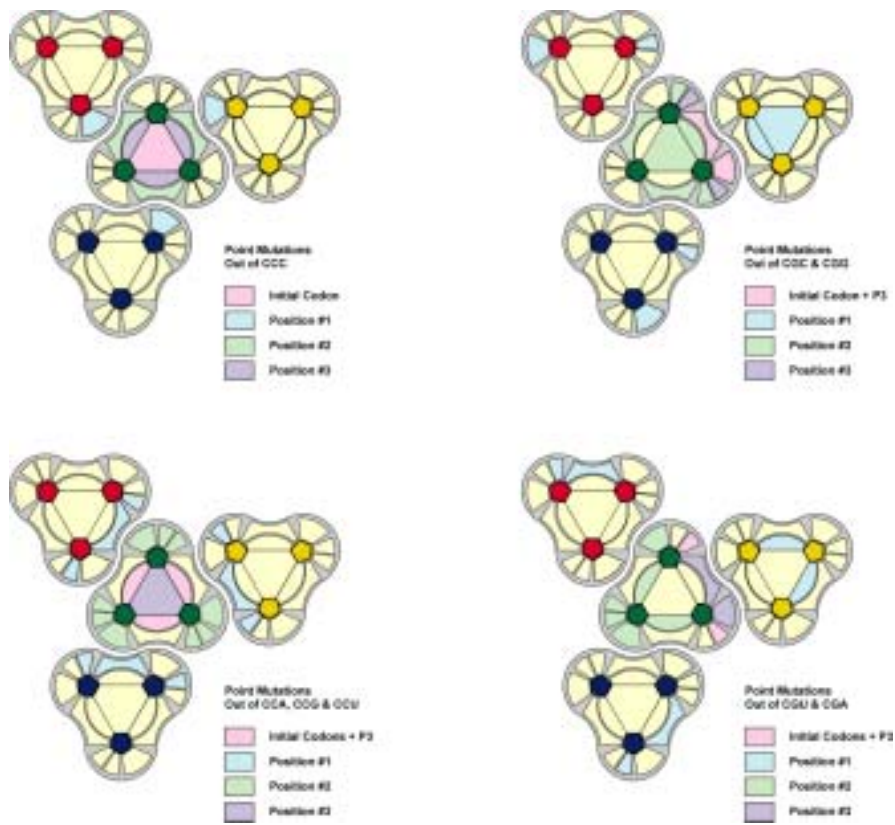

The pattern of point mutations is generally incoherent across the entire map because the various point mutations land in all codon types from all four poles, and this is merely the pattern from two multiplets! However, it still represents only a partial randomization because only a single nucleotide is changed and not all three. The only logical method to accommodate this incredibly diffuse pattern within the data is to build some type of global symmetry within the entire data set. There is no way to anticipate which nucleotide in which position in which codon a point mutation will strike, so the entire structure must somehow be prepared for global randomization. Therefore, only an assignment pattern taking account of all symmetries in the codon group could anticipate this randomizing pattern. Codon similarity at the hands of all possible point mutations is merely a manifestation of global codon symmetry.

Convincing evidence shows that the standard arrangement of amino acids does in fact minimize the effect of any point mutation on many different levels of potential "meaning" in amino acids[16]. In other words, of all the possible arrangements of this set of amino acids, nature has somehow found virtually "the best" arrangement toward minimizing the effects of point mutations. This means that the genetic code operates as a type of Gray code with respect to the effects of point mutations and their amino acid substitutions[17]. It is a global collection of "minimum steps" with respect to enacting codon change. This can only be achieved by a global symmetry pattern of amino acid assignments with respect to all individual nucleotides. This is, in fact, merely one more form of codon symmetry. All of the components must fit into a larger pattern for this trick to actually work. Codon assignments, with respect to the impact of point mutations, therefore, are yet one more example of how global symmetry has organized the genetic code.

We have now seen that with respect to whole sequence transformations there is a remarkable amount of complex symmetry within the global pattern of codon assignments. We have traditionally seen sequence transformations as an unpleasant reality that is avoided by life when possible. However, in this context we might now actually perceive them as a positive goal of the system. Transformations must occur for the system to be what it is, and the system has worked diligently to ensure that transformations occur in a logical and consistent fashion. Life makes good use of the inherent symmetry of the system at every opportunity. In some respects, the symmetry is the system. We can further confirm this observation by returning to the basic structure of our Cayley triangles and perform a quick symmetry check on each one.

Figure 31.

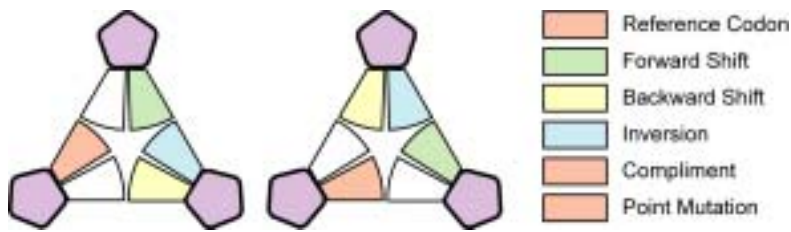

This graph represents a global symmetry key for each triangle. Two keys are presented here because of the roto-isomers between type III1 and III2 codons. However, either one of the mirror graphs can be used for any of the other codon types. The symmetry key shows us that for any specific codon, the other codons in its permutation set do a credible job of anticipating the impact of any transformation of that codon in every possible context. This is remarkably true even though randomization is always involved in these sequence transformations. For instance, the r-transformation, or a forward frameshift could produce one of four new codons, one of which will be in the actual triangle. However, because the properties of amino acids are symmetrically assigned across the global pattern, we have a very good approximation of the other three possible codons by knowing the assignment of just one. Likewise, the $r^2$-transformation, or a backward shift, does the same. Therefore, wobble groups have been assigned and then woven together to form a globally coherent pattern. The inverse permutations are literally present in the triangle. Point mutations most closely mimic the original amino acid, to the extent possible, and complements mirror the properties of their complementary codon assignments. When taken as a whole, it is a remarkable piece of symmetry work by nature. It rivals any magic square or sudoku puzzle ever conceived by man.

Each triangle acts as an informative holographic representation of the whole. The symmetry of each triangle projects itself onto the symmetry of the global pattern. However, Dr. Gamow's model, had it been correct, would have nature performing quite poorly in this exercise. That is why I call his model a proposed symmetry minimum, whereas nature apparently sought a symmetry maximum. Nature broke the symmetry of every possible codon, but it did so in the most symmetrical way possible. Just as the symmetry of codons is perfectly broken by DNA it is also perfectly broken by amino acids. This is as it should be. Every codon's symmetry is broken within the global context of the symmetry of all possible codons. So, let's now take a closer look at Dr. Gamows model of codons and his predicted assignment pattern. We can use his simple predictions of the assignment pattern to glean some insight into the actual symmetry of the codon assignments. Gamow thereby unwittingly provided us with an additional simple test for the global symmetry of amino acid assignments based on individual codons and nucleotide permutation triangles. He essentially predicted a simple yet perfectly incorrect global pattern of assignments based on these system elements. They can be seen as eighty-one individual tests of codons and triangles (ignoring stop codons and complementary triangles, as he did). Here are the criteria for testing Gamow's model:

For a **triangle to pass** it must be assigned:

- a single amino acid. **AND**
- an amino acid not in another triangle.

For a **codon to pass** it must:

- be in a passing triangle. **OR**
- share a triangle with any "synonymous" codon.

This is a generous interpretation for the compact triangle model, yet it still fails all eighty-one tests. It is never easy to propose a model that is either perfectly right or perfectly wrong. There must be some quality in the pattern of amino acid assignments that allows it to perfectly dodge Dr. Gamow's prediction. Again, it is symmetry that allows this to happen. We can appreciate this phenomenon better by first realizing that Dr. Gamow proposed that each triangle would contain an absolute minimum of assignment information. It reflects nothing more than a simple compression scheme of global assignments. This is not surprising because Dr. Gamow was merely trying to logically explain the apparently extreme level of compression in the data itself. He was searching for logical redundancy because he believed that the genetic code should be maximally redundant. He believed that codons mean amino acids, but they do not and it is not a maximally redundant code. Protein synthesis, when viewed globally, is in all probability a fabulous decompression algorithm. How could it not be? Information is maximally compressed in a genome, but we have yet to identify the complex nature of that information and therefore have failed to recognize the basic pattern of this decompression scheme. The real pattern has been tragically inverted in our minds. The genetic code in reality has the complex task of making proteins as an entire set, and it therefore must also use many "epigenetic" algorithms to eventually decompress the information into protein structures through time. This only makes sense when we realize that structure determines sequence. It only makes sense when we realize that the symmetry principle cannot be violated in nature.

In Gamow's particular graphic scheme each Cayley triangle is built upon six bits of nucleotide information but contains only 4.2 bits with respect to amino acids. However, we can quickly show that nature found a scheme that takes this set of amino acids and maximizes the information in each triangle despite the standard erroneous view of "redundancy" or compression within the genetic code. Each triangle is maximally uncompressed with respect to amino acid assignments. Symmetry was the obvious, perhaps the only tool available to do so most efficiently. In this sense, the most symmetrical arrangement is also the least redundant and the most informative with respect to amino acids. This is just as it should be.

Until now, we have focused here almost exclusively on sequence symmetry within the genetic code. However, sequencing amino acids is merely one important component in a much more complex process of synthesizing protein structures, and there are surely many different forms of symmetry involved in the larger process. After all, the genetic code can only make sequences of amino acids by first making sequences of peptide bonds. These bonds must have elements of spatial and temporal information when they are made, and surely there is symmetry in this information. Unfortunately, we know very little about the information in these bonds today. However, most functions of the genetic code are purely structural functions. All functions are performed with molecular combinations and dictated by the structural nature of molecules in general, so this is where the missing information must be hiding; it must be heretofore hiding from us in the structural relationships of the participating molecules. It must be hiding in codon combinations. To be sure, there must be other important combinations of molecular structures, starting with combinations of codons and immediately extending into combinations of anticodons, tRNA and peptide bonds. What are they? We simply do not know. In other words, the genetic code might appear to be brutally redundant with respect to simple sequence information alone, but it is not an entirely redundant system in general. In order to fully understand and appreciate the beautiful symmetry of the genetic code, we must expand our view of protein synthesis. We must make our view consistent with the symmetry principle that we rightly believe to be operating within the universe in general. We must allow our view of molecular information to include all forms of molecular information, whatever they may be. The formation of even a single protein involves a complex algorithm operating within a complex array of molecules that must interact in terms of time, space and number. Here again, the symmetry of codons is a primary tool to allow them to organize through time into a coherent array capable of doing so. It is beyond the scope of this treatment to speculate on the essential combinations of codons and their resulting maps to logical images of complex molecular sets, but it is strongly implied from this treatment that these maps in nature are founded upon the inherent symmetries of the system as a whole.

It is the r-symmetry of codons that first paves the way for wobble to become a molecular reality and a powerful tool of protein translation. The numbers of anticodons and the physical molecules that carry them can thereby be varied greatly within different molecular arrays. The concentrations of these and other molecules within the array can be varied and thereby impact the time components of translation. So, in addition to sequence and spatial symmetries, we must include an appreciation for time symmetries within the translation system. These are merely additional forms of the same symmetries within the system, and they have apparently been utilized to make the system as efficient, effective and flexible as possible across a wide diversity of organisms. Life really seems to like this arrangement, so there is a broad symmetry involved in life's formation of this entire system. In other words, life has chosen a single system with multiple, subtle variations that actually do compete[18]. Life is not now a manifestation of different symmetry systems competing for dominance. It has chosen a single system and competition is heated within that system.

Admitting the existence of global symmetry in the genetic code and the central role it has played in the pattern of codon assignments forces upon us the widely ignored issue of spatial symmetry within the system of translation. In other words, what specific role does structural symmetry play in translation and how does it play it? We have here already seen that the structure of the language itself shares the geometric symmetry of dodecahedrons and tetrahedrons, but is this geometry shared by other molecular elements of the system? A cursory examination of the translation machinery, the inputs, the intermediates and the outputs, demonstrates that, in fact, it does and we can crudely find it. Beginning with the roughly ten-fold transverse rotational symmetry of the double helix we can find the presence of the universal metric of geometric scale, which is called phi, or the golden ratio. This geometric proportion is roughly shared by the general dimensions of a standard tRNA molecule as well. An idealized form of a tRNA molecule can therefore be fitted on each codon in the G-ball and perfectly extend the spatial symmetry of this molecular apparatus into a much larger representation of the same icosidodecahedral spatial symmetry.

Figure 32.

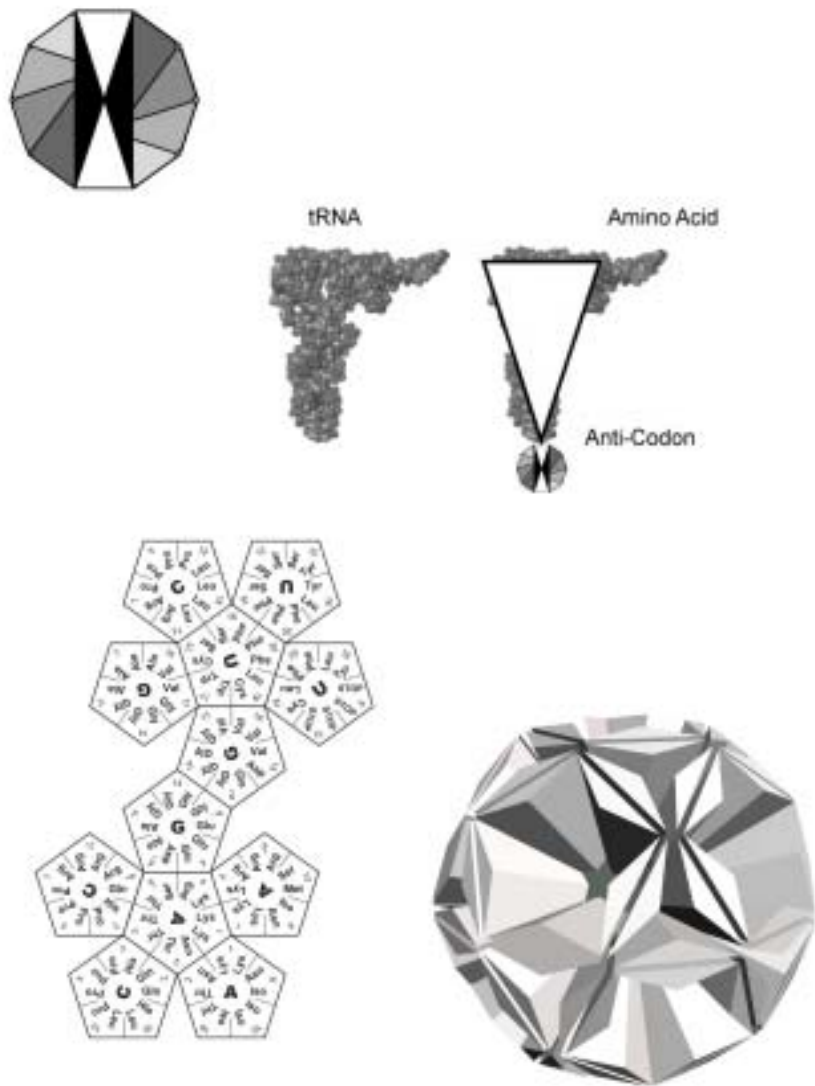

This kind of spatial symmetry can also crudely but widely be found in larger structures that rely on symmetrical sub-units to self-assemble consistently into larger organic structures, like virus particles and radiolarians.

Figure 33[19].

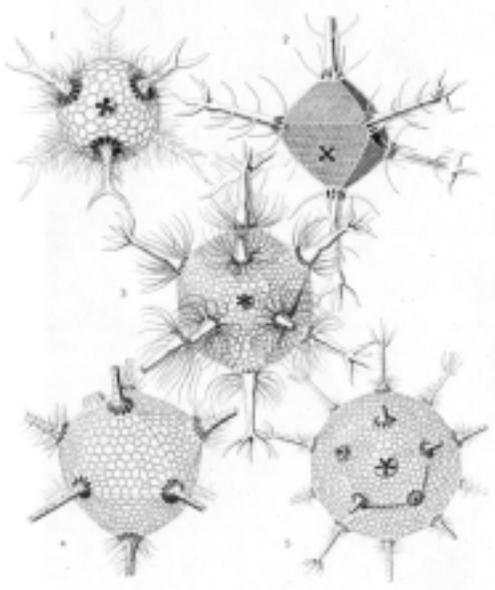

Symmetry and complex geometry are found in all systems of self-assembly in nature[20,21,22]. However, the formation of all protein structures is also a self-assembly process, and it too would greatly benefit from a coherent spatial symmetry within all of the molecules of the complex arrays of molecules involved in their formation[23]. In fact, proteins have been shown to fall into broad groups that are defined by their basic spatial symmetries generally matching the symmetries of perfect solids[24]. This means that all protein structures are broadly consistent with each other and broadly consistent with perfect structures of geometry, and therefore broadly consistent with the components of the mechanism of their own formation. It is a crystal system of crystal logic on scales too complex to yet properly comprehend.

Therefore, starting with DNA, mRNA and tRNA we can see that the genetic code is at least founded on a consistent platform of spatial geometry. Execution of the code itself is a natural exercise in molecular space filling, a purely geometric process, and the sequence symmetry of codons is also founded on dodecahedral symmetry.  It is, therefore, quite reasonable to conclude that the entire code, all of its complex dimensions of molecular information in time, space and molecular sequence, is primarily organized over vast periods of time and large numbers of molecules by the invariant laws of symmetry in the universe.  Symmetry is the primary tool for the universe to build not just inorganic molecular systems but more complex organic systems as well.  There is symmetry between the rules of self-assembly in organic and inorganic molecular systems.  A natural bridge exists for molecules to pass from inorganic to organic systems of behavior.  It is a complex process of making molecular structures fit logically together through time.  After all, this is the only mechanism available to all molecules.  This kind of abstract thinking might begin to bring the first principles of molecular biology more in line with those of sub-particle physics on which all molecular behavior is ultimately governed.

Regarding all standard simple questions about the codon table in general, "symmetry" seems to now be the universally correct answer.  In other words, symmetry correctly answers any simple question about why codons and the standard codon assignment patterns are the way they are.  Why are all of the amino acids of the l-form?  Because if they were not, they would destroy the symmetry required in translating the many sequence transformations that will naturally occur through time.  There surely were both amino acid forms available when the system got its start, but a system that excludes one form can maintain more symmetry than a system that includes both.  Why are there twenty amino acids?  Again, symmetry is the answer.  A study done in 1998 demonstrates that the sets of "synonymous" codons conform to the supersymmetry patterns of Lie algebras[25], a finding which is quite remarkable when carefully considered.  This basically means that the size of the set of amino acids, apportioned the specific way they are, represents another clear example of the symmetry groups within the codon assignment matrix.  A different number of amino acids in the matrix risks disturbing the overall symmetry of the sets and their specific assignments. After all, there are good ways and bad ways of breaking symmetry with respect to the information that will result from any breaking of symmetry. The genetic code has apparently found a "perfect" way to break codon symmetry with respect to any particular set of amino acids.

When the codon table was first elucidated it seemed somewhat of a disappointment for its apparent simplicity and lack of any obvious global patterns[26]. There seemed to be very little organizing principle behind it, so the forces that led to its organization remained obscure. It was then brazenly and erroneously called arbitrary and considered a "frozen accident," the result of a proposed selective force known as the functional imperative. In other words, once formed, the code will only function if it is executed in one way - a clever but false premise. This now absurd perception in retrospect was mostly a human artifact of the arbitrary organizational structure of the table itself. We merely failed to comprehend the nature of the codon table and what it could tell us about the genetic code. In reality, the entire pattern is neither frozen nor accidental. There is no real imperative but there can be a functional optimum[27].

There certainly is no lack of interesting patterns to behold now in the context of multi-dimensional information and codon symmetry, and these patterns are far from simple. The symmetry that defines these patterns not only ensures their subtlety but also guides their utility. The existence of so much symmetry within the translation system is hard to deny now, so we must begin to seriously question the role that symmetry has played in the formation of the system itself. In other words, did the system arise and then find its symmetry or did the inherent symmetry provide the initial engine and ultimate target for the system to form in the first place? Do we find these sets of molecules in nature because the inherent symmetry of the system found them or is the symmetry that we see merely a curious byproduct of this particular set of molecules? Quite simply, do the molecules build the G-ball or did the ball select the molecules? After all, the symmetry and logic existed before the molecules, and this set of molecules appears to merely take perfect advantage of it. It seems obvious to me that the system evolved from and toward this symmetry, and it is now strongly organized in complex ways around it. It did not start this way, it had no chance to be this complex and this precise at its origin, but it has become this way now, and there is little room for improvement under the current set of circumstances. There can be no protein without DNA and no DNA without protein, but there can be a basic form of logic between the two before either exists. The genetic code is highly functional now in executing that logic, but how functional was it everywhere along its path in getting here? It is clearly optimized in many and various ways, but what role does optimization play in past, present and future evolution of the system?

One potential answer can be found by asking an equally valid question regarding the specific role played by codon symmetry. What practical benefit is served to life by adopting this symmetrical pattern of amino acid assignments? We can only understand the answer when we understand the proper relationship between codons and proteins. Codons do not make proteins but sequences of codons do. Every sequence of codons in some way means a specific protein. However, life is not tasked merely with making single proteins in an information vacuum. It must make all proteins relative to all possible proteins and all specific combinations of actual proteins. It can only make the proteins it can make, and it can only make all sets of actual proteins relative to this limited set. Therefore, it is not enough for life to merely make proteins; it must also find acceptable proteins and compatible populations of proteins to make. It behooves life then to select from a set of self-compatible proteins. It is in this context that the value of sequence symmetry becomes apparent. When life finds a specific sequence of codons to make a useful protein, it now has a rapid way to find many other compatible proteins for it to also make. By using existing molecular sequences and natural genomic symmetries, life can efficiently search for new proteins and new protein populations. In this way the genetic code should not be seen merely as an operating system for making single proteins. The genetic code has symmetrically integrated a search engine into the code of protein synthesis. It constantly searches for new proteins and effective new sets of proteins. This is a competitive advantage for any organism trying to solve complex and ever changing problems via the application of complex protein solutions. A symmetrical genetic code provides a symmetrical set of proteins and a clear evolutionary advantage to those organisms that employ it.

The symmetry of genomes is tightly integrated into the systems of their translations. We should expect then to find the telltale fingerprints of symmetry throughout all genomes. This has proven to be low hanging fruit in the nascent world of bioinformatics. Genomes are loaded with symmetry patterns that start with the symmetry of complementary strands and extend into the palindromes of entire genomes. There is even a global symmetry fingerprint that can be found across phyla that is reflected in the patterns of codon usage and codon bias[28,29,30]. This pattern, it seems to me, is merely a reflection of the symmetry of translation multiplied over countless generations of symmetrical sequence transformations of existing sequences. It is a measure of system efficiency achieved through the minimization and symmetrical reuse of common components. Life is always efficient, and life is always a stunningly symmetric system of molecular information. This surely must include every imaginable form of molecular symmetry. This is what drives life forward in its march to accumulate molecular information at an accelerating pace, and to perfect the systems of molecular translation.

One fact is hard to deny: the entire system of protein translation considered broadly is fairly universal across all organisms over the entire planet. It took hundreds of millions of years, perhaps billions of years for this system to evolve to the level of complexity, precision and seeming ubiquity that we see today. This seems to me to logically represent an optimization of a system rather than an arbitrary arrangement that is mysteriously prevented from evolving.

The logic and symmetry of the universe provides a common magnet for convergent evolution of molecular codes. In other words, this kind of fundamental system symmetry might represent a forced move in the path of evolution. Granted, the "functional imperative" is a factor that will always retard evolution with respect to code changes, but it is not enough of a factor to ever explain the origin of this code, and clearly it is not enough of a factor to explain the many similar characteristics now seen across all life on earth in light of their obvious known differences. These codes clearly evolve, so what kept them from evolving widely apart? What drove them to exist in the first place? It appears that symmetry is again the correct general answer. But keep in mind that the standard codon table only shows relationships between codons and amino acids, and codons do not mean only amino acids in nature. Although the codon-amino acid relationship may stay relatively constant, the code itself might not. Do not allow the obvious similarities prevent you from recognizing the many differences. There are intermediates, like anticodons and tRNA, that are always selected between codons and amino acids. There is no logical upper limit to the number of these intermediates. There are also many variations above the level of amino acid sequences. Nature allows for additional flexibility at these levels because the code is not just executed in one dimension, and so life has chosen an impressive range of variations at these complex levels as well. The basic components are consistent but the complex combinations of them might not be so consistent. So, the genetic code, when taken in its entirety, is perhaps not as universal as we like to believe.[31,32] There are more variations of it than we realize, but the primary logic of any variation is based on a common symmetry, and it maintains a common logic. It can all start and it can all stay together with the perfect symmetry represented by a dodecahedron. Life expands its options from there.

There should always be strong time symmetry to our perception of evolution on earth. We should see the same processes and patterns operating to evolve our organic molecular systems today as we imagine were initially bootstrapping the system at its origins. It is, perhaps, the symmetry implied by Occam's razor. We should, therefore, not so easily accept an unnecessary complication and an ugly time asymmetry to the process, where life is allowed to evolve rapidly and wildly only up to a point and then some undefined universal policeman prohibits any further evolution. Likewise, we should not postulate a sudden, universal appearance for the code because there are clear and powerfully useful aspects to having a system where all of the organisms use roughly the same translation system. This explains the strikingly simple similarities within a broader and more complex system. By using common basic components, a constant generation and exchange of more complex components and building blocks can occur within the system. There is an obvious symmetric exchange and recombination of genetic information through time. In addition to vertical exchange, the system even widely benefits from a lateral exchange of genetic information, but only if a common system is used. A common system will, therefore, naturally accelerate its own information content and complexity whereas a fragmented system will not. Once again, the apparent universality of the genetic code is but one more example of how nature leverages symmetry in the self-assembly of organic systems of molecular information.

It is now a valid question to ask: do we have sixty-four codons simply because we start with four nucleotides in DNA or did nature arrive at these four nucleotides specifically because sixty-four codons represent an optimum within the symmetry of codons? Why not choose just two nucleotides, like a computer's binary system? After all, the geometric language between dodecahedrons and tetrahedrons as well as the sequence symmetry of complementary triplets exists prior to and independent of any molecular system of translation. A rudimentary system of molecular information could use it with any given number of nucleotides and still be remotely functional. However, this specific set of molecules - a set of dual binary nucleotides which is now also cleverly related to a compatable spectrum of amino acids - is absolutely required to reveal the beautifully intricate symmetry that we just illustrated. Any number of nucleotides will function within this system to some degree, but four mutually complementary nucleotides is the most efficient way to utilize the overall inherent symmetry of codons. How long would it take for nature to find such a thing? Once found, what would be the impetus for an organism to significantly diverge from it? What system for building intricate molecular structures could we now imagine that would be any better than the one nature is showing us today? All of these questions can only be asked and potentially answered from the context of perfect symmetry, the kind that we can most easily see from a perfectly symmetrical arrangement of the components of the system.

## What is a Codon Table?

A codon table might be many things depending on what one believes it to be. To some, a codon table is the entire genetic code. However, this limited view requires that the entire code be embodied by a simple relationship between codons and amino acids, and this view has been empirically proven false. The code is in reality a complex translation of molecular information, and when we talk of molecular information we must talk of many molecules, many sets of molecules, and the important relationships within and between sets of molecules, whatever they may be. To know the genetic code, therefore, we must first know the sets of molecules involved.

The broadest sets of molecules involved in the genetic code, the most complex sets with the most general relationships, are sets of genomes and proteins. Of course, there are many informative subsets within and between these two. Genomes are composed of nucleotides, and proteins are composed of amino acids, so naturally we are interested in the relationships between nucleotides and amino acids. To be sure, nucleotides and amino acids are the primary players, but they are not the only players. There are higher combinations of these sets, and there are many different kinds of relationships within even these sets. The relationship between codons and amino acids is but one of many, but they are not the only two sets involved, and this cannot be the only relationship of interest if we want to truly understand the entire code. So, when we look at a codon mapping with amino acids alone, we can never imagine that we are looking at the genetic code per se. We can know, however, that we are looking through a tiny and informative window into the genetic code. Some windows provide a clearer view than others, but even the clearest window of this limited sort will take an extremely agile mind to comprehend exactly what is being seen through it and what is not.

It stands to reason that we must first know a codon before we can know a codon table. A codon is not real but abstract, and so a codon table must also be abstract. We have defined a codon abstractly here as an ordered set of three nucleotides. Because it is an ordered set of any three elements, and these elements must at some point be reordered, a codon has a natural symmetry, and this is true of every codon independent of the specific identities of the elements in any particular set. However, in nature the elements of codons are themselves selected from another set of four elements, and this set also seems to be well ordered and symmetrically related. It too has symmetry that is reflected in the natural symmetry of this particular dual binary set of nucleotides. Therefore, these particular codons also inherit the symmetry brought to them by this set of nucleotides. Codons are inherently symmetrical objects, and a codon table can be a symmetrical object too. But one cannot truly understand a codon until one understands that it is at first a relationship between nucleotides, and nucleotides have a relationship between each other. Therefore, any useful map of codons should hope to start with a mapping of the relationships between nucleotides and then attempt to map the relationships between every set of three of those nucleotides. After all, a codon is only a codon within a specific context of nucleotides and within a more general context of all other codons. Meaning is derived by comparison and symmetry is perhaps the most robust form of comparison there is.

A codon only exists in nature within a specific context of nucleotide relationships, and the specific relationships between nucleotides are always informative to the codon. In other words, there are no free-floating codons in nature. Codons are not real molecules and cannot be found independently anywhere in nature. Just as no single codon has any meaning, neither does any single amino acid. In this way, all of these molecules are analogous to our system of numbers, where all numbers are abstract and all numbers derive their meaning from the relationship they have to other numbers. All codons exist only as part of longer nucleotide sequences, and the meaning of all codons is derived from those sequences. Meaning of codons is derived from relationships with other codons just as meaning of nucleotides is derived from relationships with other nucleotides and meaning of numbers is derived from the relationship to other numbers. Just as numbers can be counted and can be used to measure things, codons are counted in nature and are the measure of meaning in every sequence of nucleotides. Unlike numbers on a line, however, codons exist within a symmetrical sphere of relationships. The number line for codons is not really a line but a sphere. Nature always measures molecules in units of symmetry.

Only when we have a map of nucleotides and their relationship to codons can we adequately begin to analyze a map between codons and something else. That something else today is typically a set of amino acids, but in the future it will also need to include other things if we are ever to have a proper illustration of the entire genetic code. All maps of codons should start with the relationships between nucleotides and codons. Furthermore, we must accept the fact that no single map between codons and amino acids will ever comprehensively describe the genetic code to us.

Within this general context, it seems then that the real nature of the larger problem of "assigning" amino acids is to first assign them to nucleotides and not merely to codons. After all, nature seems to have done just that. Amino acids are not assigned statically to three nucleotides in a single context but to individual nucleotides and all possible sets of them within every possible context of all nucleotides. All nucleotides are assigned to many amino acids and all amino acids are assigned to many nucleotides. There is a global pattern to this relationship and it is not a simple one. Codons themselves are not static but dynamic. Codons are never isolated they always appear as whole sequences. They are merely relationships between nucleotides, and all nucleotides in nature are dynamic. In other words, every nucleotide in every sequence has a past present and future. Every nucleotide in every sequence today is the result of past nucleotides and will play a role in future nucleotides. The nucleotides and the sequences are always changing, so the relationships between them must constantly adjust to this level of change. Plus, the act of translation itself is always a dynamic process. Molecular sets always develop physical relationships in both time and space, and when we speak of molecular translation - that literally is molecular translation, a dynamic relationship between molecular sets in time and space. The amount of time that it takes for these relationships to actualize is a fundamental part of the system of translation on every scale of the system. Time is always a fundamental element of translation. Time is an element of codons and their assignments to amino acids. It took time to assign them and their assignments reflect the time element of change within and between them. Although any codon map can only be a single snapshot, it must always be seen as snapshot of a dynamic process that can only be fully appreciated in a much larger context.

To help visualize this, imagine a single sequence of nucleotides hanging vertically in space like a string of lights. The codon table tells us a part of its relationship to protein, but it does not tell us the whole story. Imagine that this sequence becomes coated with all of the tRNA with all of the amino acids it encodes in a particular reference frame. Imagine that this particular sequence is of a "real" protein and therefore has the inherent sequential properties of a protein. As a protein imagine that it lights up the amino acid sequence. If it were purely random, not a real protein, chances are it would not be too terribly protein-like, and therefore it would stay dark.

Imagine now that we simultaneously coat this same strand with the tRNA and amino acids from the forward shift and the backward shift, as well as the complement and its shifts. Imagine the inverse and many different combinations of transformations also coat it. Imagine some sequences with handfuls of point mutations. Imagine this process until the nucleotide strand is covered solidly with say 20-30 sequences of related tRNA and amino acids. This larger pattern of assignments is also a fundamental reflection of the pattern of the genetic code. Because all of these sequences are logically related, many of these strands will also light up as being protein-like. Now further imagine a sea made of the set of all possible nucleotide sequences, and they all exist as coated strands just like this one. Imagine how they might be logically related to each other, and imagine a global pattern of dark and light strands in an entire sea. Actual organisms will represent light patterns in this sea. It is the job of the genetic code to help efficiently find those patterns. Imagine how difficult this would be if the genetic code were not so symmetrical. After all, the assignments are not re-made after transformations, they are all made before all transformations of nucleotide sequences occur, but all of the transformations are guaranteed to occur at some point in time. A comprehensive map of the relationships between nucleotides and nucleotides, codons and codons, amino acids and amino acids, not to mention all of the tRNA, and a logical mapping of all of them to each other should give us some insight into the broader workings of the system. Symmetry is the foundation of these relationships, so our map should reflect those symmetry relationships.

## Choosing a Proper Icon for the Genetic Code

The codon table is not the genetic code but it is our visual icon for the genetic code. It immediately brings to mind that which we know, and it now must remind us of that which we do not yet know. In reality, codons mean anticodons, and anticodons mean tRNA, and tRNA mean peptide bonds, and peptide bonds mean secondary structures, and secondary structures mean tertiary structures, and tertiary structures mean quaternary structures or proteins. Proteins mean a complex population of molecules that ultimately mean life, but proteins must eventually loop back and mean DNA, RNA and more proteins. The same information that is translated into molecular structures must be able to recognize the sequential structures from which it came. The language of life is highly recursive and so it is quite complex in this way. Symmetry is the glue that binds it all together. There are many dimensions of information and they all must be logically related. It is so complex that we could never hope to capture it in a single table, but our table can still be an informative icon toward our thoughts.

The standard codon table represents a more-or-less arbitrary arrangement of a specific small set of data. There are an infinite number of ways that this data might be arranged, and all of them might in some way be informative, but none of them can be comprehensive. This means that we must choose among them. The symmetrical arrangement of the G-ball presented above is still an arrangement of this same data, but it is far less arbitrary and far more robust. It has useful structural and logical features that the standard table does not. It maps more than codons and amino acids and therefore it can greatly inform our thinking about the data and the natural processes that use it. The G-ball subsumes the table, but the converse cannot be said.

The icon that we ultimately choose to represent this natural phenomenon of molecular translation will depend on our specific needs and a direct comparison between our actual options. The linear codon table is undeniably convenient if all one needs is a simple lookup table, and if all one wants is to derive the linear sequence of amino acids in a protein in a single context. The G-ball perhaps competes poorly on terms of two-dimensional graphical convenience alone. But a side-by-side analysis based on all relevant features shows the G-ball to be superior on all counts.

**Objectivity**

The standard codon table is merely a data object, but it is conceptually a "linear" object that demonstrates an arbitrary arrangement subjectively chosen from a large number of logically equivalent structures. We might "line up" all codons in any old way that all equal the limited epistemic value of this arrangement. Therefore, the patterns observed in the data itself will always be largely subjective. The G-ball, on the other hand, is a data object chosen from only two possible objects of this form (it has a mirror twin). The data patterns seen here are un-weighted and therefore are the natural patterns of this form. The one chosen above reveals more aesthetically pleasing patterns in the data than does its mirror, so this degree of subjectivity should not be too disturbing. A dodecahedron is a real object, and it is shown here via Cayley's theorem that its symmetry elements can be used to create an isomorphic data object to completely represent the sequence symmetry of this specific set of nucleotides and codons. The G-ball wins out on comparisons of objectivity.

**Dimensionality**

The standard codon table is designed to represent and relay a single dimension of molecular information in the genetic code. That dimension consists entirely of a simple relationship between sixty-four codons and twenty amino acids. It is a time-independent and structure-independent relationship. It is merely drawn from a single instance of every sequence of three nucleotides independent of all other sequences. The genetic code, on the other hand, does not operate in this way and could never have self-organized from this first principle alone. The existence and translation of molecular information is a time-dependent process on many different levels. The synthesis of specific protein structures is now known to involve many different dimensions of molecular information always playing out on a backdrop of time. It should be clear by now that a protein structure consists of much more molecular information than merely its specific amino acid sequence, or simply its linear composition. These concepts are lost in and consistently confused by the simplistic one-dimensional structure of the standard codon table. This structure insists that we see this data strictly as having only one dimension of meaning.

Although the G-ball cannot yet be used as a secret decoder ring for other dimensions of the actual genetic code, it does illustrate the reality that more than one dimension of molecular information is involved in molecular relationships, and this additional information is somehow transmitted through time via the genetic code itself. The necessary time element is suggested by the easy transformations that can symmetrically be done on every single codon. A dimension of spatial symmetry is suggested by the physical structure itself. The dimension of shared symmetry between nucleotides, amino acids, codons and anticodons can also be conceptualized and visualized by this structure. The G-ball extends the symmetry and the relationships one level bellow codons to the level of individual nucleotides. We start here with an ordered set of twelve nucleotides not four. Every nucleotide in the G-ball is distinguishable from every other by its identity and the identities of all other nucleotides within the global structure. Any three "synonymous" nucleotides can be differentiated by their McNeil subscripts, which in this case merely tell us the identity of the nucleotide on the opposite side of the map. Unlike the DNA octahedron, the McNeil subscripts in the G-ball do not tell us the complements of nucleotides, they tell us which other nucleotide will never be found in any codon that uses that nucleotide. It is not a reference to position but to global nucleotide relationships. Each of the twelve distinct nucleotides on the map participates symmetrically in the same number and type of codons, but none of them participates in codons containing all three of the remaining nucleotide types.  This is important information in recognizing which codons will become other codons based on individual nucleotides during transformations.  It thereby shows us how amino acid assignments are further arranged around the symmetry of transformations.

The G-ball shows us dimensions of information that exist between individual nucleotides, codons and amino acids.  It is a truly multi-dimensional map of this information in many important respects.  The G-ball wins out in terms of dimensionality.

**Compression**

The standard codon table represents a partial symbolic compression of the genetic code.  It is a linear configuration merely broken down and arranged in two spatial dimensions for the sole purpose of conveniently eliminating nucleotide symbols.  It is simply a matter of graphic convenience.  Note that any truly linear arrangement of sixty-four codons would require 192 nucleotide symbols, but the table typically only employs twenty-four.  In comparison, the G-ball employs only twelve nucleotide symbols, and this is the minimum required to represent all codons in this particular set.  Therefore, the G-ball is, in fact, a maximum symbolic compression of this particular data.  There are many positive consequences to achieving this level of symbolic compression.  First, it demonstrates that the translation system can indeed be maximally compressed. Molecular information now appears to be stored in a compressed format in DNA and expanded through time in many ways by various ingenious algorithms, the genetic code perhaps being just one of them.  This perfect symbolic compression reflects the fact that there is a "most-efficient" way to organize the system in general.  Second, it demonstrates via McNeil subscripts that there are actually twelve distinct nucleotide symbols involved in the translation system.  The significance of this can only be appreciated when one admits the influence of time and sequence context on the system.  This means that any nucleotide in a specific sequence can be seen to participate through time and sequence transformations with a specific set of amino acid assignments.  This might be used to weight entire sequences of nucleotides within the context of each nucleotide to all others.

These kinds of weighting strategies will probably prove to be more effective toward pattern detection schemes when all sequence data is taken as a whole. The compressed symbolic format of the G-ball immediately tells us what the global assignments are and how each nucleotide is different in this broader context. After all, it is not a single line of assignments but a network of assignments between nucleotides, codon position and amino acids, and it ultimately produces a logical matrix of codon assignment patterns that can best be appreciated in this format. The G-ball wins out in terms of compression.

**Geometry**

The standard codon table admits of no geometry in the genetic code. Geometry is the language we use to communicate universal spatial relationships. It is a human artifact of natural phenomena. However, we have learned that when extra dimensions are required in our models of natural systems, the geometries we use must somehow be reconfigured to reflect these extra dimensions and their respective symmetries. General relativity and its dependence on non-Euclidian geometry is but one example of expanded symmetry within a geometric system.

The fundamental structure of the genetic code is built upon natural symmetry too, but this can be represented by many different geometric features of the code. Sequence symmetry is a major element of this structure, and it is a happy coincidence that we can use common geometry to illustrate this form of symmetry in the genetic code. However, the genetic code at bottom is a molecular algorithm for building molecular structures, and this is a decidedly space-filling task. Therefore, we should expect there to be elements of space filling geometry somehow involved in the algorithm as well. We presently do not know what they are, but we can easily see that all of the component molecules of this algorithm somehow share elements of a single geometric system, and this system is represented by the highest geometric spatial symmetry of a dodecahedron. As Plato said, the dodecahedron is the cosmos. It embroiders the heavens, and this is true of the universe of organic molecular information as well. The consistency between sequence symmetry and spatial symmetry within this complex molecular system is, therefore, quite striking. The G-ball captures this geometric consistency like no other icon ever could, and, therefore, it wins hands down on this comparison.

**Symmetry**

The standard codon table is an asymmetric data structure that arranges its data in an asymmetric way. This is a poor way to visualize and communicate any notion of symmetry. At the same time, the genetic code is primarily organized by symmetry, but this feature of it is always missed in the standard table. Consequently, the genetic code has been erroneously perceived in many different and flawed ways. It has been described as a fundamentally arbitrary "accident" that is somehow "frozen" in time. After all, without an organizing first principle for the genetic code, it is hard to imagine or explain the fact that the genetic code is fairly ubiquitous today, and apparently it has been more or less this way for perhaps billions of years. A far more enlightened view of molecular information, however, does not suffer from this explanatory quandary. The languages of life are primarily organized by a universal invariance that is generally known as symmetry, so a perceived invariance in this particular code merely reflects the fact that life has found a remarkably efficient way to take full advantage of those symmetries. There is no better way to communicate this concept than with a perfectly symmetric icon for the genetic code. Res ipsa loquitor. The G-ball immediately tells us that the genetic code is symmetric, and all linear arrangements tell us it is asymmetric. The G-ball, therefore, smokes the competition in terms of symmetry.

**Decision Time**

A small handful of chemists triumphantly announced to the world in the late 1960s that the genetic code had been cracked. Worse than a false alarm, this marked the starting point of a false direction. These chemists were operating under a false concept of a molecular code and with a false model of molecular translation. They had no chance of understanding and truly cracking the genetic code in this context, and there is no hope of doing so now with any single table of simple data. The genetic code has similarly not been cracked here; it has merely been re-uncracked. However, as exciting as the initial announcement surely was, it is equally exciting to now learn that there is still so much important work yet to be done. Rather than a handful of chemists, this new work will require the intensely collaborative effort of countless people that will surely span the entire spectrum of all human intellectual pursuits. This is a tremendously complex task, and we will surely benefit from a proper map and icon to guide our thinking.

The codon table is our undeniable icon for the genetic code, but all codon tables are merely abstract maps. Unfortunately, the standard table is a brutally concrete manifestation of what should be an otherwise abstract concept. The G-ball provides a new and welcome abstraction here. We find good use for maps of all kinds, but we should never confuse any map with the thing it represents in reality. Early earth explorers conceived of flat maps of our world. They imagined that there were orthogonal boundaries beyond which sea monsters lurked. The early explorers of molecular biology have been similarly duped by a flat map of an unknown world, but they are now indeed being consumed by the ideological monsters that lurk beyond its imaginary boundaries. Just as a globe is informative to our exploration of earth, a perfectly symmetrical map of nucleotides and proteins might safely guide our future explorations. We should carefully consider the options as we set sail anew into the complex and uncharted waters of molecular information.

Richard Courant wrote in, *What is Mathematics?*[33]:

"A similar situation, even more accentuated, exists is mathematics. Throughout the ages mathematicians have considered their objects, such as numbers, points, etc., as substantial things in themselves. Since these entities had always defied attempts at an adequate description, it slowly dawned on the mathematicians of the nineteenth century that the question of meaning of these objects as substantial things does not make sense within mathematics, if at all. The only relevant assertions concerning them do not refer to substantial reality; they state only the interrelations between mathematically "undefined objects" and the rules governing operations with them. What points, lines, numbers "actually" *are* cannot and need not be discussed in mathematical science. What matters and what corresponds to "verifiable" fact is structure and relationship, that two points determine a line, that numbers combine according to certain rules to form other numbers, etc. A clear insight into the necessity of a dissubstantiation of elementary mathematical concepts has been one of the most important and fruitful results of the modern postulational development.

Fortunately, creative minds forget dogmatic philosophical beliefs whenever adherence to them would impede constructive achievement. For scholars and layman alike it is not philosophy but active experience in mathematics itself that alone can answer the question: What is mathematics?"

What is true of mathematics is also true of the genetic code. I have been accused in the past of practicing numerology here. I plead guilty as charged. I believe that the numbers are important because they reveal the logic, structure and operation of the genetic code to us. Life has created a natural language of crystals, and that language is highly mathematical because it is built upon sets of crystals and the pure logic that exists between those sets. The language is complex, rivaling any language known to man because it is built of many complex sets, operating in many different environments over vast periods of time. As numbers are to mathematics molecules are to life. Molecules are the numbers in the math being performed by life. They derive their meaning only by logical comparison to other molecules. Unfortunately, no simple numbers and no simple maps of them will ever fully enlighten us to the complex beauty of this natural language, but it is the proper place for us to start. We must now strive to find the various and complex combinations of molecular sets and nature's algebra that gives them meaning.

So, a decision must be made once again as to a preferred icon. The codon table is merely a historical artifact. It is a scab serving as a painful reminder of one-half century of wounds suffered as a result of false dogma and ideological confusion. It is an icon that communicates mistakes made not lessons learned. No map of codons to amino acids can encompass the genetic code, but the icon we chose should communicate the features and structure behind this fabulous natural molecular phenomenon. To each his own, I suppose. If one merely wants a convenient way to represent a single dimension of molecular information, then it is truly hard to beat the standard codon table. However, if one is interested in communicating the essential features of a much larger concept, if one is interested in contemplating and understanding the more complex logic and structure behind a larger and seemingly incomprehensibly complex natural phenomenon, then there is no better way to start than with the G-ball. It is superior by all rational comparisons to any standard codon table. A perfectly symmetric icon is the clear and coherent path to a more enlightened and useful model of molecular information and genetic translation.

Contact: Mark White, MD  mark@codefun.com

**References**

[1] Title suggested by Subhash Kak

[2] Watson, J. & Crick, F. Molecular Structure of Deoxypentose Nucleic Acids. *Nature* 737-738 (1953).

[3] Wagner, Roy  Symbols That Stand for Themselves. The University of Chicago Press 1986

[4] Watson, Baker, Bell, Gann, Levine, Losick  Molecular Biology of the Gene. 461-478 (2004).

[5] Cortazzo P, Cerveñansky C, Marín M, Reiss C, Ehrlich R, Deana A. Silent mutations affect in vivo protein folding in Escherichia coli. Biochemical and Biophysical Research Communications. Volume 293, Issue 1, 26 April 2002, Pages 537-541

[6] Kimchi-Sarfaty C, Oh JM, Kim I, Sauna Z, Calcagno AM, Ambudkar S, Gottesman M. A "silent" polymorphism in the MDR1 gene changes substrate specificity. Science. (2007).

[7] Rosen, Joe  Symmetry Discovered: Concepts and Applications in Nature and Science.  Dover  Pages 108-124, (but read the whole thing) (1998)

[8] Livio, Mario  The Equation That Couldn't Be Solved:  How Mathematical Genius Discovered the Language of Symmetry.  Simon and Schuster  (2005)

[9] Armstrong M. A.  Groups and Symmetry.  Springer  (1988)

[10] Sands Donald D.  Introduction to Crystallography.  Dover  (1975)

[11] Hurlbut, C., Klein, Cornelis, (Kase)  Manual of Mineralogy 19th Edition.  John Wiley & Sons  (1977)

[12] Mike McNeil invented the concept of a McNeil subscript.

[13] Gamow, G., "Possible relation between deoxyribonucleic acid and protein structure.," *Nature,* vol. 173, pp. 318, (1954)

[14] Horton, Moran, Ochs, Rawn, Scrimgeour.  Principles of Biochemistry, Third Edition.  Page 59 (2002)

[15] Concept introduced to me by Richard Merrick.

[16] S.J. Freeland, T. Wu and N. Keulmann, The case for an Error Minimizing Standard Genetic Code Origins of Life and Evolution of the Biosphere 33: 457-477.

[17] D. Bosnacki, H.M.M. ten Eikelder, P.A.J. Hilbers, Genetic Code as a Gray Code Revisited.  in the Proceedings of The 2003 International Conference on Mathematics and Engineering Techniques in Medicine and Biological Sciences METMBS'03: June 23-26 2003, Las Vegas, Nevada, United States,  (2003)

[18] Darwin, Charles  The Origin of Species.  Random House (1993)

[19] Haekel, E.  Skeletons of various radiolarians.

[20] Ghyka, Matila  The Geometry of Art and Life.  Dover  (1977)

[21] Thompson, D'Arcy Wentworth  On Growth and Form  Dover revised edition (1992)

[22] Ball, Philip  The Self-Made Tapestry: Pattern Formation in Nature.  Oxford University Press  (1999)

[23] Branden, C.  Tooze, J  Introduction to Protein Structure,  Second Edition.  Garland Publishing  (1999)

[24] Denton, M. & Marshall, C.  Laws of form Revisited.  Nature Vol. 410, pg 417 (2001)

[25] Bashford, J. D. & Jarvis, P.D.  The genetic code as a periodic table: algebraic aspects.  http://www.physics.adelaide.edu.au/~jbashfor/ (2000)

[26] Hayes, Brian  The Invention of the Genetic Code.  American Scientist  Volume: 86 Number: 1 Page: 8 (1998)

[27] Freeland, S.J.  The genetic code: an adaptation for adapting? Journal of Genetic Programming and Evolvable Machines 3(2): 113-127. (2002)

[28] Gorban, A. N., Popova, T., Zinovyev, A. Y.  Four basic symmetry types in the universal 7-cluster structure of microbial genomic sequences *In Silico* Biology 5, 0025 (2005); ©2005, Bioinformation Systems e.V.

[29] Oresic, M. & Shalloway, D.  Specific Correlations between Relative Synonymous Codon Usage and Protein Secondary Structure. *Journal of Molecular Biology*, Vol. 281, No. 1, 31-48 (1998)

[30] Gupta, S., Majumdar, S., Bhattacharya, T., Ghosh, T.  Studies on the Relationships between the Synonymous Codon Usage and Protein Secondary Structural Units. *Biochemical and Biophysical Research Communications*  Vol. 269, No. 3, 692-696 (2000)

[31] Osawa, Syozo  Evolution of the Genetic Code.  Oxford University Press  (1995)

[32] Pouplana, Lluis Ribas de  The Genetic Code and the Origin of Life.  Landes Bioscience (2004)

[33] Courant, R., Robbins, H., Revised by Stewart, I.  What is Mathematics?  An Elementary Approach to Ideas and Methods.  Oxford University Press  (1996)